%% file: paper.tex
\documentclass[a4paper,11pt]{article}

\RequirePackage{lineno}

\usepackage{jheppub}
\usepackage{units}
\usepackage{xspace}
\usepackage{mathrsfs}
\usepackage{units}
\usepackage{amssymb}
\usepackage{placeins}
\usepackage{subfigure}
\usepackage{chngcntr}
\usepackage{booktabs}
\usepackage{rotating}
\usepackage{multirow}
\counterwithout{equation}{section}

\newcommand{\Alpgen}{{\textsc{Alpgen}}\xspace}

\newcommand{\Mcatnlo}{{\textsc{MC@NLO}}\xspace}
\newcommand{\aMcAtNlo}{a{\textsc{MC@NLO}}\xspace}
\newcommand{\AcerMC}{{\textsc{AcerMC}}\xspace}
\newcommand{\Sherpa}{{\textsc{Sherpa}}\xspace}
\newcommand{\Herwig}{{\textsc{Herwig}}\xspace}
\newcommand{\Herwigpp}{{\textsc{Herwig{\tt++}}}\xspace}

\newcommand{\Prospino}{{\textsc{Prospino2}}\xspace}
\newcommand{\Madgraph}{{\textsc{MadGraph}}\xspace}

\newcommand{\CTEQSixL}{{\textsc{CTEQ6L1}}\xspace}

\newcommand{\CTTen}{{\textsc{CT10}}\xspace}
\newcommand{\PYTHIA}{{\textsc{Pythia}}\xspace}
\newcommand{\GEANT}{{\textsc{GEANT4}}\xspace}

\newcommand{\MCFM}{{\textsc{MCFM}}\xspace}

\newcommand{\POWHEG}{{\textsc{Powheg}}\xspace}
\newcommand{\Perugia}{{\textsc{Perugia2011C}}\xspace}
\newcommand{\AUET}{{\textsc{AUET2B}}\xspace}
\newcommand{\AU}{{\textsc{AU2}}\xspace}
\newcommand{\ggVV}{{\text{gg2VV}}\xspace}

\newcommand{\ninoone}{{\ensuremath{\mathchoice%
      {\displaystyle\raise.4ex\hbox{$\displaystyle\tilde\chi^0_1$}}%
         {\textstyle\raise.4ex\hbox{$\textstyle\tilde\chi^0_1$}}%
       {\scriptstyle\raise.3ex\hbox{$\scriptstyle\tilde\chi^0_1$}}%
 {\scriptscriptstyle\raise.3ex\hbox{$\scriptscriptstyle\tilde\chi^0_1$}}}}}
\newcommand{\ninotwo}{{\ensuremath{\mathchoice%
      {\displaystyle\raise.4ex\hbox{$\displaystyle\tilde\chi^0_2$}}%
         {\textstyle\raise.4ex\hbox{$\textstyle\tilde\chi^0_2$}}%
       {\scriptstyle\raise.3ex\hbox{$\scriptstyle\tilde\chi^0_2$}}%
 {\scriptscriptstyle\raise.3ex\hbox{$\scriptscriptstyle\tilde\chi^0_2$}}}}}
\newcommand{\ninothree}{{\ensuremath{\mathchoice%
      {\displaystyle\raise.4ex\hbox{$\displaystyle\tilde\chi^0_3$}}%
         {\textstyle\raise.4ex\hbox{$\textstyle\tilde\chi^0_3$}}%
       {\scriptstyle\raise.3ex\hbox{$\scriptstyle\tilde\chi^0_3$}}%
 {\scriptscriptstyle\raise.3ex\hbox{$\scriptscriptstyle\tilde\chi^0_3$}}}}}
\newcommand{\ninoonetwothreefour}{{\ensuremath{\mathchoice%
      {\displaystyle\raise.4ex\hbox{$\displaystyle\tilde\chi^0_{1,2,3,4}$}}%
         {\textstyle\raise.4ex\hbox{$\textstyle\tilde\chi^0_{1,2,3,4}$}}%
       {\scriptstyle\raise.3ex\hbox{$\scriptstyle\tilde\chi^0_{1,2,3,4}$}}%
 {\scriptscriptstyle\raise.3ex\hbox{$\scriptscriptstyle\tilde\chi^0_{1,2,3,4}$}}}}}
\newcommand{\ninotwothreefour}{{\ensuremath{\mathchoice%
      {\displaystyle\raise.4ex\hbox{$\displaystyle\tilde\chi^0_{2,3,4}$}}%
         {\textstyle\raise.4ex\hbox{$\textstyle\tilde\chi^0_{2,3,4}$}}%
       {\scriptstyle\raise.3ex\hbox{$\scriptstyle\tilde\chi^0_{2,3,4}$}}%
 {\scriptscriptstyle\raise.3ex\hbox{$\scriptscriptstyle\tilde\chi^0_{2,3,4}$}}}}}
\newcommand{\nino}[1]{{\ensuremath{\mathchoice%
      {\displaystyle\raise.4ex\hbox{$\displaystyle\tilde\chi^0_{#1}$}}%
         {\textstyle\raise.4ex\hbox{$\textstyle\tilde\chi^0_{#1}$}}%
       {\scriptstyle\raise.3ex\hbox{$\scriptstyle\tilde\chi^0_{#1}$}}%
 {\scriptscriptstyle\raise.3ex\hbox{$\scriptscriptstyle\tilde\chi^0_{#1}$}}}}}
\newcommand{\chinoonep}{{\ensuremath{\mathchoice%
      {\displaystyle\raise.4ex\hbox{$\displaystyle\tilde\chi^+_1$}}%
         {\textstyle\raise.4ex\hbox{$\textstyle\tilde\chi^+_1$}}%
       {\scriptstyle\raise.3ex\hbox{$\scriptstyle\tilde\chi^+_1$}}%
 {\scriptscriptstyle\raise.3ex\hbox{$\scriptscriptstyle\tilde\chi^+_1$}}}}}
\newcommand{\chinoonem}{{\ensuremath{\mathchoice%
      {\displaystyle\raise.4ex\hbox{$\displaystyle\tilde\chi^-_1$}}%
         {\textstyle\raise.4ex\hbox{$\textstyle\tilde\chi^-_1$}}%
       {\scriptstyle\raise.3ex\hbox{$\scriptstyle\tilde\chi^-_1$}}%
 {\scriptscriptstyle\raise.3ex\hbox{$\scriptscriptstyle\tilde\chi^-_1$}}}}}
\newcommand{\chinoonepm}{{\ensuremath{\mathchoice%
      {\displaystyle\raise.4ex\hbox{$\displaystyle\tilde\chi^\pm_1$}}%
         {\textstyle\raise.4ex\hbox{$\textstyle\tilde\chi^\pm_1$}}%
       {\scriptstyle\raise.3ex\hbox{$\scriptstyle\tilde\chi^\pm_1$}}%
 {\scriptscriptstyle\raise.3ex\hbox{$\scriptscriptstyle\tilde\chi^\pm_1$}}}}}
\newcommand{\chinopmonetwo}{{\ensuremath{\mathchoice%
      {\displaystyle\raise.4ex\hbox{$\displaystyle\tilde\chi^\pm_{1,2}$}}%
         {\textstyle\raise.4ex\hbox{$\textstyle\tilde\chi^\pm_{1,2}$}}%
       {\scriptstyle\raise.3ex\hbox{$\scriptstyle\tilde\chi^\pm_{1,2}$}}%
 {\scriptscriptstyle\raise.3ex\hbox{$\scriptscriptstyle\tilde\chi^\pm_{1,2}$}}}}}
\newcommand{\chinomponetwo}{{\ensuremath{\mathchoice%
      {\displaystyle\raise.4ex\hbox{$\displaystyle\tilde\chi^\mp_{1,2}$}}%
         {\textstyle\raise.4ex\hbox{$\textstyle\tilde\chi^\mp_{1,2}$}}%
       {\scriptstyle\raise.3ex\hbox{$\scriptstyle\tilde\chi^\mp_{1,2}$}}%
 {\scriptscriptstyle\raise.3ex\hbox{$\scriptscriptstyle\tilde\chi^\mp_{1,2}$}}}}}
\newcommand{\chinopm}[1]{{\ensuremath{\mathchoice%
      {\displaystyle\raise.4ex\hbox{$\displaystyle\tilde\chi^\pm_{#1}$}}%
         {\textstyle\raise.4ex\hbox{$\textstyle\tilde\chi^\pm_{#1}$}}%
       {\scriptstyle\raise.3ex\hbox{$\scriptstyle\tilde\chi^\pm_{#1}$}}%
 {\scriptscriptstyle\raise.3ex\hbox{$\scriptscriptstyle\tilde\chi^\pm_{#1}$}}}}}
\newcommand{\chinomp}[1]{{\ensuremath{\mathchoice%
      {\displaystyle\raise.4ex\hbox{$\displaystyle\tilde\chi^\mp_{#1}$}}%
         {\textstyle\raise.4ex\hbox{$\textstyle\tilde\chi^\mp_{#1}$}}%
       {\scriptstyle\raise.3ex\hbox{$\scriptstyle\tilde\chi^\mp_{#1}$}}%
 {\scriptscriptstyle\raise.3ex\hbox{$\scriptscriptstyle\tilde\chi^\mp_{#1}$}}}}}
\newcommand{\stau}{\ensuremath{\tilde{\tau}}\xspace}
\newcommand{\slep}{\ensuremath{\tilde{\ell}}\xspace}

\newcommand{\stauL}{\ensuremath{\tilde{\tau}_{L}}\xspace}
\newcommand{\slepL}{\ensuremath{\tilde{\ell}_{L}}\xspace}

\newcommand{\stauR}{\ensuremath{\tilde{\tau}_{R}}\xspace}
\newcommand{\slepR}{\ensuremath{\tilde{\ell}_{R}}\xspace}
\newcommand{\snu}{\ensuremath{\tilde{\nu}}\xspace}
\newcommand{\snut}{\ensuremath{\tilde{\nu}_{\tau}}\xspace}
\def\slepton{\ensuremath{\tilde{\ell}}}
\def\sleptonR{\ensuremath{\tilde{\ell}_{\mathrm{R}}}} 
\newcommand{\ttbar}{\ensuremath{t\bar{t}}\xspace}

\newcommand{\pt}{\ensuremath{p_\mathrm{T}}}
\newcommand{\mt}{\ensuremath{m_\mathrm{T}}}
\newcommand{\msfos}{\ensuremath{m_\mathrm{SFOS}}}
\newcommand{\mttwo}{\ensuremath{m_\mathrm{T2}^\mathrm{max}}}
\newcommand{\mlt}{\ensuremath{m_{\ell\tau}}}
\newcommand{\mtt}{\ensuremath{m_{\tau\tau}}}
\newcommand{\mlll}{\ensuremath{m_{3\ell}}}
\newcommand{\mindPhi}{\ensuremath{\Delta\phi_{\ell\ell'}^\mathrm{min}}}
\newcommand{\pTvec}{\vec{p}_\mathrm{T}}
\newcommand{\qTvec}{\vec{q}_\mathrm{T}}

\newcommand{\pTell}[1]{\vec{p}_\mathrm{T,{\, #1}}}
\newcommand{\pTmiss}{\vec{p}_\mathrm{T}^\mathrm{\,miss}}
\newcommand{\met}{{\ensuremath{E_{\mathrm{T}}^{\mathrm{miss}}}}}
\newcommand{\ifb}{\mbox{fb$^{-1}$}}
\newcommand{\lumi}{\unit[20.3]{\ifb}\xspace}

\def\TeV{\ifmmode {\mathrm{\ Te\kern -0.1em V}}\else
                   \textrm{Te\kern -0.1em V}\fi}%
\def\GeV{\ifmmode {\mathrm{\ Ge\kern -0.1em V}}\else
                   \textrm{Ge\kern -0.1em V}\fi}%
\def\MeV{\ifmmode {\mathrm{\ Me\kern -0.1em V}}\else
                   \textrm{Me\kern -0.1em V}\fi}%
\def\keV{\ifmmode {\mathrm{\ ke\kern -0.1em V}}\else
                   \textrm{ke\kern -0.1em V}\fi}%
\def\eV{\ifmmode  {\mathrm{\ e\kern -0.1em V}}\else
                   \textrm{e\kern -0.1em V}\fi}%

\usepackage{preprintcover}  
\PreprintCoverPaperTitle{\boldmath Search for direct production of charginos and neutralinos in events with three leptons and missing transverse momentum in $\sqrt{s}\,=\,$8$\TeV$ $pp$ collisions with the ATLAS detector}  
\PreprintIdNumber{CERN-PH-EP-2014-019}  
\PreprintCoverAbstract{A search for the direct production of charginos and neutralinos in final states with three leptons and missing transverse momentum is presented. 
The analysis is based on \lumi of $\sqrt{s}$ = 8$\TeV$ proton--proton collision data delivered by the Large Hadron Collider and recorded with the ATLAS detector.
Observations are consistent with the Standard Model expectations and limits are set in $R$-parity-conserving phenomenological Minimal Supersymmetric Standard Models and in simplified supersymmetric models, significantly extending previous results. 
For simplified supersymmetric models of direct chargino ($\chinoonepm$) and next-to-lightest neutralino ($\ninotwo$) production with decays to lightest neutralino ($\ninoone$) via either all three generations of sleptons, staus only, gauge bosons, or Higgs bosons, $\chinoonepm$ and $\ninotwo$ masses are excluded up to $700 \GeV$, $380 \GeV$, $345 \GeV$, or $148 \GeV$ respectively, for a massless $\ninoone$.
}  
\PreprintJournalName{JHEP}  

\begin{document}


 \title{\boldmath Search for direct production of charginos and neutralinos in events with three leptons and missing transverse momentum in $\sqrt{s}\,=\,$8$\TeV$ $pp$ collisions with the ATLAS detector}

\author{The ATLAS Collaboration}

\abstract{ 
A search for the direct production of charginos and neutralinos in final states with three leptons and missing transverse momentum is presented. 
The analysis is based on \lumi of $\sqrt{s}$ = 8$\TeV$ proton--proton collision data delivered by the Large Hadron Collider and recorded with the ATLAS detector.
Observations are consistent with the Standard Model expectations and limits are set in $R$-parity-conserving phenomenological Minimal Supersymmetric Standard Models and in simplified supersymmetric models, significantly extending previous results. 
For simplified supersymmetric models of direct chargino ($\chinoonepm$) and next-to-lightest neutralino ($\ninotwo$) production with decays to lightest neutralino ($\ninoone$) via either all three generations of sleptons, staus only, gauge bosons, or Higgs bosons, $\chinoonepm$ and $\ninotwo$ masses are excluded up to $700 \GeV$, $380 \GeV$, $345 \GeV$, or $148 \GeV$ respectively, for a massless $\ninoone$.
}

\maketitle
\flushbottom


\section{Introduction}

Supersymmetry (SUSY)~\cite{Miyazawa:1966,Ramond:1971gb,Golfand:1971iw,Neveu:1971rx,Neveu:1971iv,Gervais:1971ji,Volkov:1973ix,Wess:1973kz,Wess:1974tw} proposes the existence of supersymmetric particles, with spin differing by one-half unit with respect to that of their Standard Model (SM) partners. 
Charginos, $\chinopmonetwo$, and neutralinos, $\ninoonetwothreefour$, collectively referred to as electroweakinos, are the ordered mass eigenstates formed from the linear superposition of the SUSY partners of the Higgs and electroweak gauge bosons (higgsinos, winos and binos).
Based on naturalness arguments~\cite{Barbieri:1987fn,deCarlos:1993yy}, the lightest electroweakinos are expected to have mass of order 100$ \GeV$ and be accessible at the Large Hadron Collider (LHC).
In the $R$-parity-conserving minimal supersymmetric extension of the SM (MSSM)~\cite{Fayet:1976et,Fayet:1977yc,Farrar:1978xj,Fayet:1979sa,Dimopoulos:1981zb}, SUSY particles are pair-produced and the lightest SUSY particle (LSP), assumed in many models to be the $\ninoone$, is stable. 
Charginos and neutralinos can decay into leptonic final states via superpartners of neutrinos ($\snu$, sneutrinos) or charged leptons ($\slepton$, sleptons), or via $W$, $Z$ or Higgs ($h$) bosons ($\chinopm{i} \rightarrow \ell^{\pm}\tilde{\nu},\nu{\tilde{\ell}}^{\pm},W^{\pm}\nino{j},Z\chinopm{j},h\chinopm{j}$ and $\nino{i} \rightarrow\nu\snu,\ell^{\pm}{\tilde{\ell}}^{\mp},W^{\pm}\chinomp{j},Z\nino{j},h\nino{j}$ respectively). 

This paper presents a search performed with the ATLAS detector for the direct production of charginos and neutralinos decaying to a final state with three charged leptons ($e$, $\mu$ or $\tau$, referred to as leptons in the following) and missing transverse momentum originating from the two undetected LSPs and the neutrinos.
The analysis is based on 20.3$\,$fb$^{-1}$ of proton-proton collision data recorded by ATLAS at a centre-of-mass energy of $\sqrt{s}\,$$=\,$8$\TeV$.
Previous searches for charginos and neutralinos are documented in refs~\cite{2012ku,:2012gg,ATLASmlep2011} by ATLAS, and in ref.~\cite{CMS:2012ew} by CMS. 
Similar searches were conducted at the Tevatron~\cite{D0-2009,CDF-2008}. At LEP~\cite{LEPSUSYWG:01-03.1,Heister:2003zk,Abdallah:2003xe,Acciarri:1999km,Abbiendi:2003sc}, searches for direct chargino production set a model-independent lower limit of 103.5$ \GeV$ at 95\% confidence level (CL) on the mass of promptly decaying charginos.

\section{SUSY scenarios \label{sec:SUSY}}

\begin{figure}[h]
\centering
\subfigure[~$\slepL$-mediated]{\includegraphics[width=0.35\textwidth]{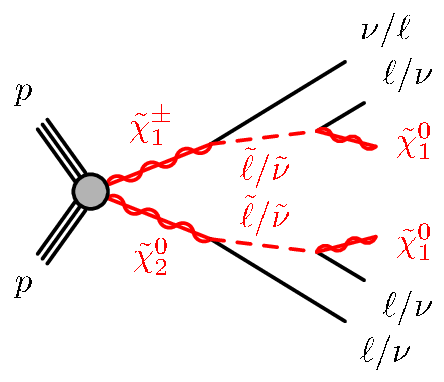}}
\hspace{2cm}
\subfigure[~$\stauL$-mediated]{\includegraphics[width=0.35\textwidth]{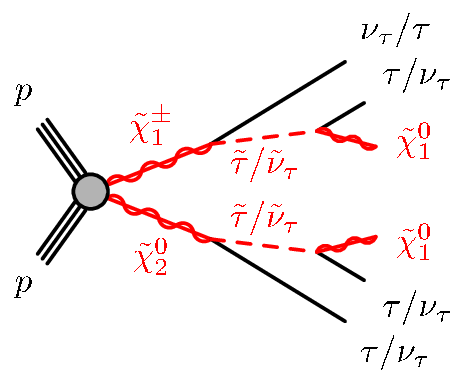}}
\subfigure[~$WZ$-mediated]{\includegraphics[width=0.35\textwidth]{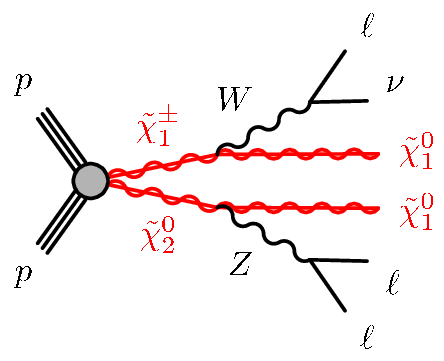}}
\hspace{2cm}
\subfigure[~$Wh$-mediated]{\includegraphics[width=0.35\textwidth]{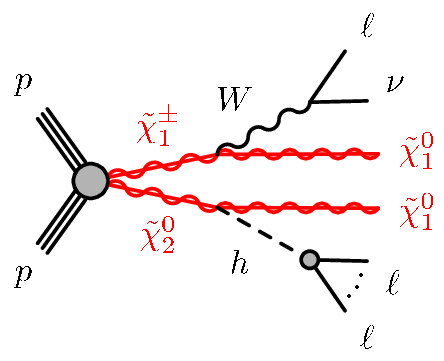}}
\caption{The Feynman diagrams for the four simplified models of the direct production of $\chinoonepm\ninotwo$ studied in this paper. 
The different decay modes are discussed in the text. 
The dots in (d) depict possible additional decay products of the lightest Higgs boson decaying via intermediate $\tau\tau$, $WW$ or $ZZ$ states.
\label{fig:Feyn} }
\end{figure}

Among the electroweakino pair-production processes leading to three leptons in the final state, $\chinoonepm \ninotwo$ production has the largest cross-section in most of the MSSM parameter space. 
Several simplified supersymmetric models (``simplified models''~\cite{Alwall:2008ag}) are considered for the optimisation of the search and interpretation of results. 
The simplified models target the direct production of $\chinoonepm$ and $\ninotwo$, where the masses and the decay modes of the relevant particles ($\chinoonepm$, $\ninoone$, $\ninotwo$, $\snu$, $\slepL$\footnote{The sleptons are referred to as left- or right-handed (e.g. $\slepL$ or $\slepR$), depending on the helicity of the fermionic superpartners.}) are the only free parameters. 
It is assumed that the $\chinoonepm$ and $\ninotwo$ consist purely of the wino component and are degenerate in mass, while the $\ninoone$ consists purely of the bino component.
Four different scenarios for the decay of the $\chinoonepm$ and $\ninotwo$ are considered, where in all cases the decays are prompt,
\begin{description}\itemsep0pt
\item[\boldmath$\slepL$-mediated:] the $\chinoonepm$ and $\ninotwo$ decay with a branching fraction of 1/6 via $\tilde{e}_L$, $\tilde{\mu}_L$, $\tilde{\tau}_L$, $\tilde{\nu}_e$, $\tilde{\nu}_\mu$, or $\snut$ with masses $m_{\snu}=m_{\slepL}\,$$=\,$$(m_{\ninoone} + m_{\chinoonepm} )/2$,
\item[\boldmath$\stauL$-mediated:] the first- and second-generation sleptons and sneutrinos are assumed to be heavy, so that the $\chinoonepm$ and $\ninotwo$ decay with a branching fraction of 1/2 via $\stau$ or $\snut$ with masses $m_{\snut}=m_{\stau}\,$$=\,$$(m_{\ninoone} + m_{\ninotwo} )/2$,
\item[\boldmath $WZ$-mediated:] all sleptons and sneutrinos are assumed to be heavy, and the $\chinoonepm$ and $\ninotwo$ decay via $W^{(*)}$ and $Z^{(*)}$ bosons, respectively, with a branching fraction of 100\%,
\item[\boldmath $Wh$-mediated:] all sleptons and sneutrinos are assumed to be heavy, and the $\chinoonepm$ and $\ninotwo$ decay via $W$ and lightest Higgs bosons, respectively, with a branching fraction of 100\%. 
The Higgs boson considered is SM-like, with a mass of 125$ \GeV$ and is assumed to decay with SM branching ratios.

\end{description}
\noindent Diagrams for the considered $\chinoonepm\,\ninotwo$ production and decay modes are shown in figure~\ref{fig:Feyn}.

Results are also interpreted in dedicated phenomenological MSSM (pMSSM)~\cite{Djouadi:1998di} scenarios, which consider all relevant SUSY production processes. 
In the models considered in this paper, the masses of the coloured sparticles, of the CP-odd Higgs boson, and of the left-handed sleptons are set to high values to allow only the direct production of charginos and neutralinos via $W$/$Z$ bosons and their decay via right-handed sleptons, gauge bosons and Higgs bosons. 
By tuning the mixing in the top-squark sector, the value of the lightest Higgs boson mass is set close to 125$\GeV$, which is consistent with the mass of the observed Higgs boson~\cite{Aad:2012tfa,Chatrchyan:2012ufa}.  
The mass hierarchy, composition and production cross-sections of the electroweakinos are governed by the ratio of the expectation values of the two Higgs doublets $\tan\beta$, the gaugino mass parameters $M_1$ and $M_2$, and the higgsino mass parameter $\mu$.
For the hierarchy $M_1\,$$<\,$$M_2\,$$<\,$$\mu$ ($M_1\,$$<\,$$\mu\,$$<\,$$M_2$), the $\ninoone$ is bino-like, the $\chinoonepm$ and $\ninotwo$ are wino-like (higgsino-like) and the dominant electroweakino production process leading to a final state with three leptons is $pp\rightarrow\chinoonepm$$\ninotwo$ ($pp\rightarrow\chinoonepm\ninotwo$, $pp\rightarrow\chinoonepm$$\ninothree$). 
If  $M_2\,$$<\,$$M_1\,$$<\,$$\mu$ ($\mu\,$$<\,$$M_1\,$$<\,$$M_2$), the $\ninoone$ ($\ninoone$, $\ninotwo$) and the $\chinoonepm$ are wino-like (higgsino-like) with similar masses and the dominant process leading to a final state with three high transverse momentum leptons is the pair-production of the higgsino-like (wino-like) $\chinopm{2}$ and the bino-like $\ninotwo$ ($\ninothree$).

Finally, the pMSSM scenarios under study are parametrised in the $\mu$--$M_2$ plane and are classified based on the masses of the right-handed sleptons into three groups,

\begin{description}\itemsep0pt
\item[pMSSM \boldmath$\slepR$:] the right-handed sleptons are degenerate in mass, with mass $m_{\tilde{\ell}_R}\,$$=\,$$(m_{\tilde{\chi}_1^{0}} + m_{\tilde{\chi}_2^{0}} )/2$. 
Setting the parameter $\tan\beta = 6$ yields comparable $\ninotwo$ branching ratios into each slepton generation.
The $\chinoonepm$ decays predominantly via a $W$ boson when kinematically allowed and to $\stau$ otherwise because the sleptons are right-handed.
To probe the sensitivity for different $\ninoone$ compositions, three values of $M_1$ are considered: 100$ \GeV$, 140$ \GeV$ and 250$ \GeV$,

\item[pMSSM \boldmath$\stauR$:] the selectrons and smuons are heavy and the $\stauR$ mass is set to  $m_{\stauR}\,$$=\,$$(m_{\tilde{\chi}_1^{0}} + m_{\tilde{\chi}_2^{0}} )/2$ and $\tan\beta\,$ to $\,$50, hence decays via right-handed staus dominate. The parameter $M_1$ is set to 75$ \GeV$  resulting in a bino-like $\ninoone$,

\item[pMSSM no \boldmath$\slep$:] all sleptons are heavy so that decays via $W$, $Z$ or Higgs bosons dominate.
The remaining parameters are $M_1\,$$=\,$50$\GeV$ and $\tan\beta\,$$=\,$10.
The Higgs branching fractions are SM-like across much of the parameter space considered. 
However, the $h \rightarrow \ninoone \ninoone$ branching fraction rises to $\sim$20\% ($\sim$70\%) when $\mu$ decreases to $200\ (100) \GeV$, suppressing other decay modes, but this does not affect the mass limits significantly.

\end{description}
%

\section{The ATLAS detector}
The ATLAS detector~\cite{atlas-det} is a multi-purpose particle physics detector with forward-backward symmetric cylindrical geometry.\footnote{ATLAS uses a right-handed coordinate
  system with its origin at the nominal interaction point (IP) in the
  centre of the detector and the $z$-axis along the beam pipe. The
  $x$-axis points from the IP to the centre of the LHC ring, and the
  $y$-axis points upward. Cylindrical coordinates $(r,\phi)$ are used
  in the transverse plane, $\phi$ being the azimuthal angle around the
  $z$-axis. The pseudorapidity is defined in terms of the polar angle
  $\theta$ as $\eta=-\ln\tan(\theta/2)$.}
 The inner tracking detector (ID) covers $|\eta|\,$$<\,$2.5 and consists of a silicon pixel detector, a semiconductor microstrip detector, and a transition radiation tracker. The ID is surrounded by a thin superconducting solenoid providing a 2$\,$T axial magnetic field. A high-granularity lead/liquid-argon (LAr) sampling calorimeter measures the energy and the position of electromagnetic showers within $|\eta|\,$$<\,$3.2. Sampling calorimeters with LAr are also used to measure hadronic showers in the end-cap (1.5$\,$$<\,$$|\eta|\,$$<\,$3.2) and forward (3.1$\,$$<\,$$|\eta|\,$$<\,$4.9) regions, while an iron/scintillator tile calorimeter measures hadronic showers in the central region ($|\eta|\,$$<\,$1.7). The muon spectrometer (MS) surrounds the calorimeters and consists of three large superconducting air-core toroid magnets, each with eight coils, a system of precision tracking chambers ($|\eta|\,$$<\,$2.7), and fast trigger chambers ($|\eta|\,$$<\,$2.4). A three-level trigger system~\cite{atlas} selects events to be recorded for offline analysis.

\section{Monte Carlo simulation}
Monte Carlo (MC) generators are used to simulate SM processes and new physics signals relevant to this analysis. 
The SM processes considered are those that can lead to leptonic signatures. 
The diboson production processes considered include $WW$, $WZ$ and $ZZ$ (where ``$Z$'' also includes virtual photons), and the $W\gamma$ and $Z\gamma$ processes. 
The triboson processes considered are $WWW$ and $ZWW$ (collectively referred to as $VVV$), while samples of SM Higgs boson production via gluon fusion, vector-boson-fusion or in association with $W$/$Z$ bosons or $\ttbar$ are also studied. 
The $\ttbar$, single top-quark, $W$+jets, $Z$+jets, $\ttbar W$, $\ttbar Z$, $\ttbar WW$, and $tZ$ processes are also considered, where $\ttbar W$, $\ttbar Z$ and $\ttbar WW$ are collectively referred to as $\ttbar V$. 
Details of the MC simulation samples used in this paper, as well as the order of cross-section calculations in perturbative QCD used for yield normalisation are shown in table~\ref{tab:MCsamples}.

\begin{table}[h]
  \centering
  \caption{For the MC samples used in this paper for background estimates, the generator type, the order of cross-section calculations used for yield normalisation, names of parameter tunes used for the underlying event generation and PDF sets. 
  \label{tab:MCsamples} }
  \tiny{
    \begin{tabular}{ ccccc }
      \toprule
Process & Generator & Cross-section  & Tune   & PDF set  \\
 & + fragmentation/hadronisation & & & \\
\midrule
{\bf Dibosons} & & & & \\
         $WW$, $WZ$, $ZZ$ & \POWHEG-r2129~\cite{Nason:2004rx,Frixione:2007vw} & NLO QCD  & \AU~\cite{Pythia8tunes} & \CTTen~\cite{CT10pdf} \\ 
 & + \PYTHIA-8.165~\cite{Sjostrand:2006za} & with \MCFM-6.2~\cite{mcfm1,mcfm2}  & & \\
         * $WZ$, $ZZ$ & \aMcAtNlo-2.0.0.beta3~\cite{Frixione:2002ik} & NLO QCD  & \AU & \CTTen \\ 
 & + \Herwig-6.520~\cite{herwig} & with \MCFM-6.2  & & \\
 & (or + \PYTHIA-6.426) &   & & \\

         $ZZ$ via gluon fusion & \ggVV~\cite{Kauer:2012hd}  & NLO & \AUET~\cite{mc11ctunes} & \CTTen \\ 
 {\it ~~~~(not incl. in \POWHEG)} &  + \Herwig-6.520 & & & \\ 
         $W\gamma$, $Z\gamma$ & \Sherpa-1.4.1~\cite{Sherpa} & NLO & (internal) & \CTTen \\  
\midrule
{\bf Tribosons} & & & & \\
        $WWW$, $ZWW$ & \Madgraph-5.0~\cite{Alwall:2007st} + \PYTHIA-6.426 & NLO~\cite{Campanario:2008yg} & \AUET & \CTEQSixL~\cite{Pumplin:2002vw} \\ 
\midrule
{\bf Higgs} & & & & \\
      via gluon fusion & \POWHEG-r2092 + \PYTHIA-8.165  & NNLL QCD, NLO EW~\cite{Dittmaier:2012vm} & \AU & \CTTen  \\ 
      via vector-boson-fusion & \POWHEG-r2092 + \PYTHIA-8.165 & NNLO QCD, NLO EW~\cite{Dittmaier:2012vm} & \AU & \CTTen  \\ 
      associated $W$/$Z$ production & \PYTHIA-8.165 & NNLO QCD, NLO EW~\cite{Dittmaier:2012vm} & \AU & \CTEQSixL \\
      associated $\ttbar$-production & \PYTHIA-8.165 & NNLO QCD~\cite{Dittmaier:2012vm} & \AU & \CTEQSixL \\
\midrule
{\bf Top+Boson} & & & & \\
          $\ttbar W$, $\ttbar Z$ & \Alpgen-2.14~\cite{Mangano:2002ea} + \Herwig-6.520 & NLO~\cite{ttZ, ttW} & \AUET & \CTEQSixL \\ 
          * $\ttbar W$, $\ttbar Z$ & \Madgraph-5.0 + \PYTHIA-6.426 &  NLO & \AUET & \CTEQSixL \\ 
          $\ttbar WW$ & \Madgraph-5.0 + \PYTHIA-6.426 &  NLO~\cite{ttW} & \AUET &  \CTEQSixL \\ 
          $tZ$ & \Madgraph-5.0 + \PYTHIA-6.426 &  NLO~\cite{tZ} & \AUET &  \CTEQSixL \\ 
\midrule
\boldmath $\ttbar$ & \POWHEG-r2129 + \PYTHIA-6.426 & NNLO+NNLL~\cite{Cacciari:2011hy,Baernreuther:2012ws,Czakon:2012zr,Czakon:2012pz,Czakon:2013goa,Czakon:2011xx} & \Perugia & \CTTen  \\ 
\midrule
{\bf Single top} & & & & \\
        $t$-channel & \AcerMC-38~\cite{Kersevan:2004yg} + \PYTHIA-6.426 & NNLO+NNLL~\cite{Kidonakis:2011wy}  & \AUET & \CTEQSixL \\  
        $s$-channel, $Wt$ &  \Mcatnlo-4.06~\cite{Frixione:2005vw,Frixione:2008yi} + \Herwig-6.520 & NNLO+NNLL~\cite{Kidonakis:2010tc, Kidonakis:2010ux} & \AUET & \CTTen \\  
\midrule
{\bf \boldmath $W$+jets, $Z$+jets} & \Alpgen-2.14 + \PYTHIA-6.426 & DYNNLO-1.1~\cite{Catani:2009sm} & \Perugia & \CTEQSixL \\ 
 & (or + \Herwig-6.520) & with MSTW2008 NNLO~\cite{Martin:2009iq} & & \\
      \bottomrule
    \end{tabular} 
  }       
\end{table}

For all MC samples, the propagation of particles through the ATLAS detector is modelled with \GEANT~\cite{Agostinelli:2002hh} using the full ATLAS detector simulation~\cite{:2010wqa}, except the $\ttbar$ \POWHEG sample, for which a fast simulation using a parametric response of the electromagnetic and hadronic calorimeters is used~\cite{atlfastII}. 
The effect of multiple proton--proton collisions from the same or nearby beam bunch crossings (in-time or out-of-time pile-up) is incorporated into the simulation by overlaying additional minimum-bias events generated with \PYTHIA onto hard-scatter events. 
Simulated events are weighted to match the distribution of the number of interactions per bunch crossing observed in data, but are otherwise reconstructed in the same manner as data. 

The SUSY signal samples  are produced with \Herwigpp-2.5.2~\cite{herwigplusplus} using the \CTEQSixL PDF set. 
Signal cross-sections are calculated to NLO in the strong coupling constant using \Prospino~\cite{Beenakker:1996ch}.
They are in agreement with the NLO calculations matched to resummation at next-to-leading logarithmic accuracy (NLO+NLL) within $\sim$2\% \cite{Fuks:2012qx,Fuks:2013vua,Fuks:2013lya}.
The nominal cross-section and the uncertainty are taken from the center and spread, respectively, of the envelope of cross-section predictions using different PDF sets and factorisation and renormalisation scales, as described in ref.~\cite{Kramer:2012bx}.

\section{Event reconstruction}
Events recorded during stable data-taking conditions are analysed if the primary vertex has five or more tracks with transverse momentum $\pt\,$$>\,$400$\MeV$ associated with it.
The primary vertex of an event is identified as the vertex with the highest $\Sigma \pt^2$ of associated tracks.
After the application of beam, detector and data-quality requirements, the total luminosity considered in this analysis corresponds to \lumi.

Electron candidates must satisfy ``medium'' identification criteria, following ref.~\cite{Aad:2011mk} (modified for 2012 data conditions), and are required to have $|\eta|\,$$<\,$2.47 and $\pt\,$$>\,$10$ \GeV$. Electron $\pt$ and $|\eta|$ are determined from the calibrated clustered energy deposits in the electromagnetic calorimeter and the matched ID track, respectively.
Muon candidates are reconstructed by combining tracks in the ID and tracks in the MS~\cite{PhysRevD.85.072004} and are required to have $|\eta|\,$$<\,$2.5 and $\pt\,$$>\,$10$ \GeV$. 
Events containing one or more muons that have transverse impact parameter with respect to the primary vertex $|d_0|\,$$>\,$0.2$\,$mm or longitudinal impact parameter with respect to the primary vertex $|z_0|\,$$>\,$1$\,$mm are rejected to suppress cosmic-ray muon background. 

Jets are reconstructed with the anti-$k_t$ algorithm~\cite{Cacciari:2008gp} with a radius parameter of $\Delta R \equiv \sqrt{(\Delta\phi)^2+(\Delta\eta)^2}\,$$=\,$0.4 using three-dimensional calorimeter energy clusters as input. 
The clusters are calibrated using the so-called local hadronic calibration, which consists of weighting differently the energy deposits arising from the electromagnetic and hadronic components of the showers~\cite{atlas}.
The final jet energy calibration corrects the calorimeter response to the true particle-level jet energy~\cite{Aad:2011he,Aad:2012vm}, where correction factors are obtained from simulation and then refined and validated using data. 
Corrections for in-time and out-of-time pile-up are also applied. 
Jets are required to have $|\eta|\,$$<\,$2.5 and $\pt\,$$>\,$20$ \GeV$ and a ``jet vertex fraction'' (JVF) larger than 0.5, if $\pt\,$$<\,$50$ \GeV$. 
The JVF is the $\pt$-weighted fraction of the tracks in the jet that are associated with the primary vertex. 
Requiring large JVF values suppresses jets from a different interaction in the same beam bunch crossing. 
Events containing jets failing to satisfy the quality criteria described in ref.~\cite{Aad:2011he} are rejected to suppress events with large calorimeter noise and non-collision backgrounds. 

Jets are identified as originating from $b$-quarks (referred to as $b$-tagged), using a multivariate technique based on quantities related to reconstructed secondary vertices. 
The chosen working point of the $b$-tagging algorithm~\cite{btag} correctly identifies $b$-quark jets in simulated $\ttbar$ samples with an efficiency of 80\%, with a light-flavour jet misidentification probability of about 4\%.

Hadronically decaying $\tau$ leptons ($\tau_{\rm had}$) are reconstructed using jets described above with $\pt\,$$>\,$10$ \GeV$ and $|\eta|\,$$<\,$2.47. 
The $\tau_{\rm had}$ reconstruction algorithm uses the electromagnetic and hadronic cluster shapes in the calorimeters, as well as information about the tracks within $\Delta R\,$$=\,$0.2 of the seed jet. 
Discriminating track and cluster variables are used within a boosted decision tree algorithm (BDT) to optimise $\tau_{\rm had}$ identification. 
Electrons misidentified as $\tau_{\rm had}$ candidates are vetoed using transition radiation and calorimeter information. 
The $\tau_{\rm had}$ candidates are corrected to the $\tau$ energy scale~\cite{tauperf} using an $\eta$- and $\pt$-dependent calibration. 
The $\tau_{\rm had}$ candidates are required to have one or three associated tracks (prongs) as $\tau$ leptons predominantly decay to either one or three charged pions, together with a neutrino and often additional neutral pions. 
The $\tau_{\rm had}$ candidates are also required to have $\pt\,$$>\,$20$ \GeV$ and have unit total charge.

The missing transverse momentum, $\pTmiss$ (and its magnitude $\met$), is the negative vector sum of the transverse momenta of all $\pt\,$$>\,$10$ \GeV$ muons, $\pt\,$$>\,$10$ \GeV$ electrons, $\pt\,$$>\,$10$ \GeV$ photons, $\pt\,$$>\,$20$ \GeV$ jets, and calibrated calorimeter energy clusters with $|\eta|\,$$<\,$4.9 not associated with these objects. 
Hadronically decaying $\tau$ leptons are included in the $\pTmiss$ calculation as jets. 
Clusters associated with electrons, photons and jets are calibrated to the scale of the corresponding objects. 
Calorimeter clusters not associated with these objects are calibrated using both calorimeter and tracker information~\cite{Aad:2012re}.
For jets, the calibration includes the pile-up correction described above, whilst the jet vertex fraction requirement is not considered when selecting jet candidates. 

In this analysis, ``tagged'' leptons are candidate leptons separated from each other and from jets in the following order:
\vspace{-5pt}
\begin{enumerate}\itemsep0pt
\item if two electron candidates are reconstructed with $\Delta R\,$$<\,$0.1, the lower energy candidate is discarded to avoid double counting. 
\item jets within $\Delta R\,$$=\,$0.2 of an electron candidate,
and $\tau_{\rm had}$ candidates within $\Delta R\,$$=\,$0.2 of an electron or muon, are rejected to avoid double counting. 
\item electron and muon candidates are rejected if found within $\Delta R\,$$=\,$0.4 of a jet to suppress semileptonic decays of $c$- and $b$-hadrons. 
\item to reject bremsstrahlung, close-by electron and muon candidates are both rejected if found within $\Delta R\,$$=\,$0.01 (0.05 for close-by muon pairs). 
\item jets found within $\Delta R\,$$=\,$0.2 of a ``signal'' $\tau$ lepton (see below) are rejected, to avoid double counting. 
\end{enumerate}
Finally, to suppress low mass resonances, if tagged electrons and muons form a same-flavour opposite-sign (SFOS) pair with $\msfos\,$$<\,$12$ \GeV$, both leptons in the pair are rejected.

Tagged leptons satisfying additional identification criteria are called ``signal'' leptons.
Signal $\tau$ leptons must satisfy ``medium'' identification criteria~\cite{tauBDT}. 
Signal electrons (muons) are tagged electrons (muons) for which the scalar sum of the transverse momenta of tracks within a cone of $\Delta R\,$$=\,$0.3 around the lepton candidate is less than 16\% (12\%) of the lepton $\pt$.
Tracks used for the electron (muon) isolation requirement defined above are those which have $\pt\,$$>\,$0.4 (1.0)$ \GeV$ and $|z_0| \,$$<\,$2$\,$mm with repect to the primary vertex of the event. 
Tracks of the leptons themselves as well as tracks closer in $z_0$ to another vertex (that is not the primary vertex) are not included. 
The isolation requirements are imposed to reduce the contributions from semileptonic decays of hadrons.
Signal electrons must also satisfy ``tight'' identification criteria~\cite{Aad:2011mk} (modified for 2012 data conditions) and the sum of the extra transverse energy deposits in the calorimeter (corrected for pile-up effects) within a cone of $\Delta R = 0.3$ around the electron candidate must be less than 18\% of the electron $\pt$.
To further suppress electrons and muons originating from secondary vertices, the $d_0$ normalised to its uncertainty is required to be small, with $|d_{0}|/\sigma(d_{0})\,<\,5\,(3)$, and $|z_{0}\sin{\theta}|\,$$<\,$0.4$\,$mm (1$\,$mm) for electrons (muons).

\section{Event selection}
\label{sec:selection}

Events are required to have exactly three tagged leptons, passing signal lepton requirements and separated from each other by $\Delta R\,$$>\,$0.3. 
At least one electron or muon is required among the three leptons. 
The signal electrons and muons in the events must have fired at least one of the single- or double-lepton triggers and satisfied the corresponding $\pt$-threshold requirements shown in table~\ref{tab:triggers}.
The $\pt$ thresholds are chosen such that the overall trigger efficiency with respect to the selected events is in excess of 90\%, and is independent of the lepton transverse momenta within uncertainties. 
The same requirements are applied to the MC-simulated events. 
Events are further required not to contain any $b$-tagged jets to suppress contributions from top-quark production. 

In the following, signal electrons and muons are labelled as  $\ell$ or $\ell'$ where the flavour of $\ell$ and $\ell'$ is assumed to be different.
Signal $\tau_{\rm had}$ are referred  to as $\tau$.
Five main signal regions are defined according to the flavour and charge of the leptons, as shown in table~\ref{tab:SRdef}, and are labelled by the number of $\tau$ leptons selected:

\begin{table}[t]
  \begin{center}
  \caption{The triggers used in the analysis and the $\pt$ threshold used, ensuring that the lepton(s) triggering the event are in the plateau region of the trigger efficiency. Muons are triggered within a restricted range of $|\eta|<2.4$.
  \label{tab:triggers} }
  \scriptsize{
    \begin{tabular}{ cc }
      \toprule
      Trigger   & $\pt$ threshold [\hspace{-2pt}$\GeV$] \\
      \midrule
      Single Isolated $e$ & 25\\
      Single Isolated $\mu$ & 25\\
      \midrule
      \multirow{2}{*}{Double $e$} & 14,14 \\
                                  & 25,10 \\
      \midrule
      \multirow{2}{*}{Double $\mu$} &  14,14 \\
                                    & 18,10 \\
      \midrule
      \multirow{2}{*}{Combined $e\mu$ } & 14($e$),10($\mu$) \\
                                        & 18($\mu$),10($e$) \\
      \bottomrule
    \end{tabular} 
  }       
  \end{center}
\end{table}

\begin{table}[b]
\centering
 \caption{
Summary of the selection requirements for the signal regions.
The index of the signal region corresponds to the number of required $\tau$ leptons.
The SR0$\tau$a bin definitions are shown in table~\ref{tab:SR0a-bins}. 
Energies, momenta and masses are given in units of $\GeV$.
The signal models targeted by the selection requirements are also shown.
\label{tab:SRdef}}
\scriptsize{
 \begin{tabular}{l c c c c c}
  \toprule
  Signal region		    &   SR0$\tau$a & SR0$\tau$b & SR1$\tau$ & SR2$\tau$a & SR2$\tau$b \\
\midrule
Flavour/sign & $\ell^{+}\ell^{-}\ell$, $\ell^{+}\ell^{-}\ell'$ & $\ell^{\pm}\ell^{\pm}\ell'^{\mp}$ & $\tau^{\pm}\ell^{\mp}\ell^{\mp}$, $\tau^{\pm}\ell^{\mp}\ell'^{\mp}$ & $\tau\tau\ell$ & $\tau^{+}\tau^{-}\ell$ \\
$b$-tagged jet & veto & veto & veto & veto & veto \\ 
$\met$ & binned & $>50$ & $>50$ & $>50$ & $>60$ \\
\midrule
Other & $\msfos$ binned  & $\pt^{ \rm{3^{rd}} \ell}\,$$>\,$20 & $\pt^{ \rm{2^{nd}} \ell}\,$$>\,$30 & $\mttwo\,$$>\,$100 & $\sum \pt^{\tau}\,$$>\,$110 \\
 & $\mt$  binned & $\mindPhi\,$$\leq\,$1.0        & $\sum \pt^{\ell}\,$$>\,$70       &               & 70$\,<\,$$\mtt\,$$<\,$120 \\
 &  &                            & $\mlt\,$$<\,$120                &               &  \\
 &  &  & $m_{ee}$ $Z$ veto &  & \\
\midrule
Target model & $\slep$,$WZ$-mediated & $Wh$-mediated & $Wh$-mediated & $\stauL$-mediated & $Wh$-mediated \\
\bottomrule
 \end{tabular}
}
\end{table}

\begin{description}\itemsep0pt
\item[\boldmath SR0$\tau$a ($\ell^{+}\ell^{-}\ell$, $\ell^{+}\ell^{-}\ell'$) --]  
a signal region composed of 20 disjoint bins defined in table~\ref{tab:SR0a-bins} is optimised for maximum sensitivity to the $\slepL$-mediated and $WZ$-mediated scenarios.
SR0$\tau$a also offers sensitivity to the $Wh$-mediated scenario. 
This signal region requires a pair of SFOS leptons among the three leptons and has five slices in $\msfos$ (defined as the invariant mass of the SFOS lepton pair closest to the $Z$ boson mass). 
Each $\msfos$ slice is further divided into four bins using $\met$ and $\mt$ selections (see table~\ref{tab:SR0a-bins}), where $\mt$ is the transverse mass formed using the $\met$ and the lepton not forming the SFOS lepton pair with mass closest to the $Z$ boson mass, $m_{\rm T}(\pTvec^{\, \ell},\pTmiss)\,$$=\,$$\sqrt{2 \pt^{\, \ell} \met - 2 \pTvec^{\, \ell} \cdot \pTmiss}$. 
Events with trilepton mass, $\mlll$, close to the $Z$ boson mass ($|\mlll -m_{Z}|\,$$<\,$10$ \GeV$) are vetoed in some bins with low $\met$ and low $\mt$ to suppress contributions from $Z$ boson decays with converted photons from final-state radiation. 
The $WZ$ and $\ttbar$ backgrounds generally dominate the SR0$\tau$a bins in varying proportions, with $WZ$ mainly dominating the bins for which $\msfos$ is in the $81.2--101.2 \GeV$ range.

\item[\boldmath SR0$\tau$b ($\ell^{\pm}\ell^{\pm}\ell'^{\mp}$) --]  
optimised for maximum sensitivity to the $Wh$-mediated scenario, this signal region vetoes SFOS lepton pairs among the three leptons to effectively suppress the $WZ$ background. 
Requirements on $\met$, lepton $\pt$ and the minimum $\Delta\phi$ between two opposite-sign (OS) leptons, $\mindPhi$, are used to reduce the backgrounds.
The remaining dominant processes are $\ttbar$ and $VVV$ production.

\item[\boldmath SR1$\tau$ ($\tau^{\pm}\ell^{\mp}\ell^{\mp}$, $\tau^{\pm}\ell^{\mp}\ell'^{\mp}$) --]  
a signal region requiring one $\tau$ and two same sign (SS) electrons or muons ($e^{\pm}e^{\pm}$, $e^{\pm}\mu^{\pm}$, $\mu^{\pm}\mu^{\pm}$) is optimised for maximum sensitivity to the $Wh$-mediated scenario.
To increase the sensitivity to the $h \rightarrow \tau \tau$ decay, $\mlt$ is required to be less than 120$\GeV$, where $\mlt$ is obtained using the $\ell$ and $\tau$ forming the pair closest to the Higgs boson mass of 125$ \GeV$.
Electron pairs with mass consistent with a $Z$ boson ($m_{ee}\,$$=\,$81.2--101.2$ \GeV$) are vetoed to suppress events in which an electron's charge is assigned the wrong sign. 
After requirements on lepton $\pt$, the diboson and $\ttbar$ processes dominate the background. 

\item[\boldmath SR2$\tau$a ($\tau\tau\ell$) --]  
this signal region is optimised for maximum sensitivity to the $\stauL$-mediated scenario and also offers some sensitivity to the $\slepL$-mediated scenario. 
It selects events with high $\met$ and high ``stransverse mass'' $\mttwo$~\cite{Lester:1999tx,Barr:2003rg}. 
The stransverse mass is calculated as $\mttwo = \min_{\qTvec}\left[\max\left(\mt(\pTell{1},\,\qTvec),\mt(\pTell{2},\,\pTmiss-\qTvec)\right)\right]$, where $\pTell{1}$ and $\pTell{2}$ are the transverse momenta of the two leptons yielding the largest stransverse mass, and $\qTvec$ is a transverse vector that minimises the larger of the two transverse masses $\mt$. 
The dominant background of this signal region is $\ttbar$ production.

\item[\boldmath SR2$\tau$b ($\tau^{+}\tau^{-}\ell$) --] 
this signal region is optimised for maximum sensitivity to the $Wh$-mediated scenario and requires two OS $\tau$ leptons to target the $h \rightarrow \tau\tau$ decay. 
Requirements on the $\pt^{\tau}$ and $\met$ provide background suppression and 
the $\tau\tau$ invariant mass $\mtt$ is required to be consistent with that resulting from a Higgs boson decay (and lower than 125$\GeV$ due to escaping neutrinos). 
Diboson and $\ttbar$ processes survive the SR2$\tau$b selection.   

\end{description}
All signal regions are disjoint, with the exception of SR2$\tau$a and SR2$\tau$b.

\begin{table}[h]
\centering
\caption{
Summary of the bins in $\msfos$, $\mt$, and $\met$ for SR0$\tau$a.
All dimensionful values are given in units of $\GeV$.
\label{tab:SR0a-bins} }
\scriptsize{
\begin{tabular}{ c r@{--}l r@{--}l  r@{--}l c c}
    \toprule
     SR0$\tau$a bin & \multicolumn{2}{c}{$\msfos$} & \multicolumn{2}{c}{$\mt$} & \multicolumn{2}{c}{$\met$} & 3$\ell$ $Z$ veto\\
    \midrule
    1 & 12 & 40              & 0 & 80              & 50 & 90 & no \\
    2 & 12 & 40              & 0 & 80              & \multicolumn{2}{c}{$>\,$90} & no \\ 
    3 & 12 & 40              & \multicolumn{2}{c}{$>\,$80}        & 50 & 75 & no \\
    4 & 12 & 40              & \multicolumn{2}{c}{$>\,$80}        & \multicolumn{2}{c}{$>\,$75} & no \\
    \midrule
    5 & 40 & 60              & 0 & 80               & 50 & 75& yes \\
    6 & 40 & 60              & 0 & 80               & \multicolumn{2}{c}{$>\,$75} & no \\
    7 & 40 & 60              & \multicolumn{2}{c}{$>\,$80}        & 50 & 135 & no \\
    8 & 40 & 60              & \multicolumn{2}{c}{$>\,$80}        & \multicolumn{2}{c}{$>\,$135} & no \\
    \midrule
    9 & 60 & 81.2            & 0 & 80               & 50 & 75& yes \\
    10 & 60 & 81.2            & \multicolumn{2}{c}{$>\,$80}        & 50 & 75& no \\
    11 & 60 & 81.2            & 0 & 110               & \multicolumn{2}{c}{$>\,$75} & no \\
    12 & 60 & 81.2            & \multicolumn{2}{c}{$>\,$110}        & \multicolumn{2}{c}{$>\,$75} & no \\
    \midrule
    13 & 81.2 & 101.2         & 0 & 110              & 50 & 90& yes \\
    14 & 81.2 & 101.2         & 0 & 110              & \multicolumn{2}{c}{$>\,$90}&  no \\
    15 & 81.2 & 101.2         & \multicolumn{2}{c}{$>\,$110}       & 50 & 135& no  \\
    16 & 81.2 & 101.2         & \multicolumn{2}{c}{$>\,$110}       & \multicolumn{2}{c}{$>\,$135}& no \\
    \midrule
    17 & \multicolumn{2}{c}{$>\,$101.2}     & 0 & 180               & 50 & 210& no \\
    18 & \multicolumn{2}{c}{$>\,$101.2}     & \multicolumn{2}{c}{$>\,$180}        & 50 & 210& no \\
    19 & \multicolumn{2}{c}{$>\,$101.2}     & 0 & 120               & \multicolumn{2}{c}{$>$\,210} & no \\
    20 & \multicolumn{2}{c}{$>\,$101.2}     & \multicolumn{2}{c}{$>$\,120}        & \multicolumn{2}{c}{$>$\,210} & no \\
    \bottomrule
\end{tabular}
}
\end{table}

\section{Standard Model background estimation}
\label{sec:background}

Several SM processes lead to events with three signal leptons. 
Lepton candidates can be classified into three main types, depending on their origin: ``real'' leptons are prompt and isolated; 
``fake'' leptons can originate from a misidentified light-flavour quark or gluon jet (referred to as ``light flavour''); 
``non-prompt'' leptons can originate from a semileptonic decay of a heavy-flavour quark, or an electron from a photon conversion. 
The SM background processes are  classified into ``irreducible'' background if they lead to events with three or more real leptons, 
or into ``reducible'' background if the event has at least one fake or non-prompt lepton. 
The predictions for irreducible and reducible backgrounds are tested in validation regions (section~\ref{sec:bgval}).

\subsection{Irreducible background processes}
Irreducible processes include diboson ($WZ$ and $ZZ$), $VVV$, $\ttbar V$, $tZ$ and Higgs boson production. 
The irreducible background contributions are determined using the corresponding MC samples, for which $b$-tagged jet selection efficiencies and misidentification probabilities, lepton efficiencies, as well as the energy and momentum measurements of leptons and jets are corrected to account for differences with respect to the data. 

\subsection{Reducible background processes \label{sec:redbg}}
Reducible processes include single- and pair-production of top quarks, $WW$ production and single $W$ or $Z$ boson produced in association with jets or photons. 
In signal regions with fewer than two $\tau$ leptons, the dominant reducible background component is $\ttbar$, followed by $Z$+jets, whereas for signal regions with two $\tau$ leptons, the dominant component is $W$+jets. 
The reducible background is estimated using a ``matrix method'' similar to that described in ref.~\cite{Aad:2010ey} and which was previously used in ref.~\cite{2012ku}. 

In this implementation of the matrix method, the highest-$p_{\rm T}$ signal electron or muon is taken to be real. 
Simulation studies show that this is a valid assumption in $>$95\% of three-signal-lepton events. 
The number of observed events with one or two fake or non-prompt leptons is then extracted from a system of linear equations relating the number of events with two additional signal or tagged leptons to the number of events with two additional candidates that are either real, fake or non-prompt. 
The coefficients of the linear equations are functions of the real-lepton identification efficiencies and of the fake and non-prompt lepton misidentification probabilities, both defined as a fraction of the corresponding tagged leptons passing the signal lepton requirements.

The real-lepton identification efficiencies are obtained from MC simulation in the signal or validation region under consideration to account for detailed kinematic dependencies and are multiplied by correction factors to account for potential differences with respect to the data. 
The correction factors are obtained from a control region rich in $Z\rightarrow e^{+}e^{-}$ and  $Z\rightarrow \mu^{+}\mu^{-}$ decays and defined with one signal and one tagged lepton, forming a SFOS pair with $|\msfos - m_Z|<10 \GeV$. 
The real-lepton efficiency correction factors are found to be $0.998\pm0.013$ and $0.996\pm0.001$ for electrons and muons respectively, where the uncertainties are statistical. 

The fake and non-prompt lepton misidentification probabilities are calculated as the weighted averages of the corrected, type- and process-dependent, misidentification probabilities defined below according to their relative contributions in a given signal or validation region.
The type- and process-dependent misidentification probabilities for each relevant fake and non-prompt lepton type (heavy flavour, light flavour or conversion) and for each reducible background process are obtained using simulated events with one signal and two tagged leptons and parameterised with the lepton $p_{\rm T}$ and $\eta$.  
These misidentification probabilities are then corrected using the ratio (``correction factor'') of the misidentification probability in data to that in simulation obtained from dedicated control samples. 
The correction factors are assumed to be independent of selected regions and any potential composition or kinematic differences. 
For non-prompt electrons and muons from heavy-flavour quark decays, the correction factor is measured in a $b\bar{b}$-dominated control sample. 
This is defined by selecting events with only one $b$-tagged jet (containing a muon candidate) and a tagged lepton, for which the misidentification probability is measured. 
Contaminating backgrounds leading to the production of real leptons from $W$ decays include top-quark pair-production and $W$ bosons produced in association with $b$-tagged jets. 
A requirement that $\met\,$$<\,$60$ \GeV$ suppresses both the $\ttbar$ and the $W$ contamination, 
and requiring $m_{\rm T}\,$$<\,$50$ \GeV$ (constructed using the tagged lepton) further reduces the $W$ background. 
The remaining ($\sim$1\% level) background is subtracted from data using MC predictions. 
The heavy-flavour correction factor is found to be $0.74 \pm 0.04$ ($0.89 \pm 0.03$) for electrons (muons), where the uncertainties are statistical. 

Fake $\tau$ leptons predominantly originate from light-flavour quark jets.
The corresponding correction factor is measured in a $W$+jets-dominated control sample, where events with one signal muon with $\pt\,$$>\,$25$ \GeV$ and one tagged $\tau$ are selected.  
The muon and $\tau$ must be well separated from all other leptons and jets in the event. 
To suppress $Z\rightarrow \tau\tau$ contributions, $\mt^\mu$$>$60$ \GeV$ and 
$\cos\Delta\phi(\pTmiss,\pTvec^{\, \mu})+\cos\Delta\phi(\pTmiss,\pTvec^{\, \tau})<-0.15$ are imposed. 
Finally, $b$-tagged jets are vetoed to suppress heavy-flavour contributions.
The light-flavour correction factor decreases from 0.9 to 0.6 (1.0 to 0.6) as the $\pt$ increases from 20$ \GeV$ to 150$ \GeV$ for one-prong (three-prong) $\tau$ decays. 

The correction factor for the conversion candidates is determined in events with a converted photon radiated from a muon in $Z\rightarrow\mu\mu$ decays. 
These are selected by requiring two oppositely charged signal muons and one tagged electron, assumed to originate from the converted photon, such that $|m_{\mu\mu e} - m_{Z}|\,$$<\,$10$\GeV$. 
The conversion correction factor for electrons is 1.14$\pm$0.12, where the uncertainty is statistical, and is independent of electron $\pt$ and $\eta$.

\subsection{Systematic uncertainties \label{sec:systematics} }
Several sources of systematic uncertainties are considered for the SM background estimates and signal yield predictions. 
The systematic uncertainties affecting the simulation-based estimates (the yield of the irreducible background, the cross-section-weighted misidentification probabilities, and the signal yield) include 
the theoretical cross-section uncertainties due to the choice of renormalisation and factorisation scales and PDFs, 
the acceptance uncertainty due to PDFs, the choice of MC generator, 
the uncertainty on the luminosity (2.8\%~\cite{Aad:2011dr}), 
the uncertainty due to the jet energy scale, jet energy resolution, 
lepton energy scale, lepton energy resolution 
and lepton identification efficiency, 
the uncertainty on the $\met$ from energy deposits not associated with reconstructed objects,
and the uncertainty due to $b$-tagging efficiency and mistag probability.
The systematic uncertainty associated with the simulation of pile-up is also taken into account. 
An uncertainty is applied to MC samples to cover differences in efficiency seen between the trigger in data and the MC trigger simulation.

The theoretical cross-section uncertainties for the irreducible backgrounds used in this analysis are 
30\% for $\ttbar V$~\cite{ttW,ttZ}, 
50\% for $tZ$, 
5\% for $ZZ$, 7\% for $WZ$ 
and 100\% for the triboson samples. 
The ATLAS $WZ$ and $ZZ$ cross-section measurements~\cite{Aad:2012twa,Aad:2012awa} are in agreement with the \MCFM predictions used here. 
For the Higgs boson samples, 20\% uncertainty is used for $VH$ and vector-boson-fusion production, 
while 100\% uncertainty is assigned to $\ttbar H$ and Higgs boson production via gluon fusion~\cite{Dittmaier:2012vm}. 
The uncertainties on $tZ$, tribosons, $\ttbar H$ and Higgs boson production via gluon fusion are assumed to be large to account for uncertainties on the acceptance, while the inclusive cross-sections are known to better precision. 
The uncertainty on the $WZ$ and $ZZ$ acceptance due to the choice of MC generator, parton showering and scales is determined by comparing estimates from \POWHEG and \aMcAtNlo, while those for $\ttbar V$ are determined by comparing \Alpgen and \Madgraph estimates.

The uncertainty on the reducible background includes the MC uncertainty on the weights for the misidentification probabilities from the sources listed in section~\ref{sec:redbg} (2--14\%) and the uncertainty due to the dependence of the misidentification probability on $\met$ (0--7\%), $\mt$ (1--7\%), $m_{\ell\ell}$ (0--18\%), SFOS selection/veto (0--5\%) and $\eta$ (1--5\%). 
Also included in the uncertainty on the reducible background is the uncertainty on the correction factors for the misidentification probability, the statistical uncertainty on the data events used to apply the matrix equation and the statistical uncertainty from the misidentification probability measured in simulation.

\subsection{Background modelling validation \label{sec:bgval}}

\begin{table}[t]
  \centering
  \caption{Summary of the selection requirements for the validation regions. 
Energies, momenta and masses are given in units of $\GeV$.
\label{tab:lepVRDef}}
  \tiny{
    \begin{tabular}{l c c c c c c c c}
      \toprule
      Region name       & N($\ell$) & N($\tau$) & Flavour/sign & $Z$ boson  & $\met$  & N($b$-tagged jets) & Target process \\
      \midrule
      VR0$\tau$noZa & 3         & 0         & $\ell^{+}\ell^{-}\ell$, $\ell^{+}\ell^{-}\ell'$  & $\msfos$ \& $m_{3\ell}$ veto & 35--50  & --     & $WZ^{*}$, $Z^{*}Z^{*}$, $Z^{*}$+jets         \\
      VR0$\tau$Za   & 3         & 0         & $\ell^{+}\ell^{-}\ell$, $\ell^{+}\ell^{-}\ell'$   & request    & 35--50        & --          & $WZ$, $Z$+jets        \\
      \midrule

      VR0$\tau$noZb & 3         & 0         & $\ell^{+}\ell^{-}\ell$, $\ell^{+}\ell^{-}\ell'$   & $\msfos$ \& $m_{3\ell}$ veto & $>$ 50  & 1  & $\ttbar$      \\
      VR0$\tau$Zb   & 3         & 0         & $\ell^{+}\ell^{-}\ell$, $\ell^{+}\ell^{-}\ell'$ & request    & $>$ 50        & 1       & $WZ$     \\
      VR0$\tau$b    & 3         & 0         & $\ell^{+}\ell^{-}\ell$, $\ell^{+}\ell^{-}\ell'$  & binned     & binned        & 1       & $WZ$, $\ttbar$ \\
      \midrule
      VR1$\tau$a & 2         & 1    & $\tau^{\pm}\ell^{\mp}\ell^{\mp}$, $\tau^{\pm}\ell^{\mp}\ell'^{\mp}$     & --      & 35--50       & --          & $WZ$, $Z$+jets \\
      VR1$\tau$b & 2         & 1    & $\tau^{\pm}\ell^{\mp}\ell^{\mp}$, $\tau^{\pm}\ell^{\mp}\ell'^{\mp}$     & --      & $>$ 50       & 1       & $\ttbar$ \\
      \midrule
      VR2$\tau$a   & 1         & 2    & $\tau\tau\ell$       & --    & 35--50       & --          & $W$+jets, $Z$+jets \\
      VR2$\tau$b   & 1         & 2    & $\tau\tau\ell$       & --    & $>$ 50       & 1       & $\ttbar$ \\
      \bottomrule
    \end{tabular}
  }
\end{table}

The background predictions are tested in validation regions that are defined to be adjacent to, yet disjoint from, the signal regions. 
For each $\tau$ multiplicity considered, validation regions are defined with either low-$\met$ (``a'' regions) or high-$\met$ + $b$-tagged jet (``b'' regions) to target different background processes. 
The definition of the regions and the targeted processes are shown in table~\ref{tab:lepVRDef}. 
In the validation region requiring no $\tau$ leptons, both the $Z$-veto and $Z$-request regions are tested in the low-$\met$ and high-$\met$ + $b$-tagged jet regions. 
To validate the binned signal region SR0$\tau$a, an orthogonal validation region (VR0$\tau$b) is defined with the same binning as shown in table~\ref{tab:SR0a-bins} and a $b$-tagged jet requirement.

\begin{table}[b]
  \begin{center}
  \caption{
Expected numbers of SM background events and observed numbers of data events in selected validation regions, as defined in table~\ref{tab:lepVRDef}. 
The binned validation region VR0$\tau$b is displayed in figure~\ref{fig:VR0b-summary}. 
Statistical and systematic uncertainties are included (as described in section~\ref{sec:systematics}).
CL$_b$ values are given.
  \label{tab:bkgLepVR}}
  \tiny{
\renewcommand\arraystretch{1.5}
    \begin{tabular}{ l   cccccccc }
\toprule
Sample   & VR0$\tau$noZa  & VR0$\tau$Za  & VR0$\tau$noZb  & VR0$\tau$Zb  & VR1$\tau$a   & VR1$\tau$b  & VR2$\tau$a  & VR2$\tau$b  \\
\midrule
 $WZ$  &  $91\pm 12$   &  $471\pm 47$   &  $10.5^{+1.8}_{-2.0}$   &  $58\pm 7$   &  $14.6\pm 1.9$   &  $1.99\pm0.35$   &  $14.3^{+2.4}_{-2.5}$   &  $1.9\pm 0.4$   \\
 $ZZ$  &  $19\pm 4$   &  $48\pm 7$   &  $0.62\pm 0.12$   &  $2.6\pm 0.4$   &  $1.76^{+0.29}_{-0.28}$   &  $0.138\pm 0.028$   &  $1.8\pm0.4$   &  $0.12\pm 0.04$   \\
 $\ttbar V$ + $tZ$  &  $3.2\pm 1.0$   &  $10.1^{+2.3}_{-2.2}$   &  $9.5\pm 3.1$   &  $18\pm 4$   &  $0.9\pm 0.9$   &  $2.8\pm 1.3$   &  $1.0\pm 0.7$   &  $1.7\pm 0.7$   \\
 $VVV$  &  $1.9\pm 1.9$   &  $0.7\pm 0.7$   &  $0.35^{+0.36}_{-0.35}$   &  $0.18\pm 0.18$   &  $0.4\pm 0.4$   &  $0.08\pm 0.08$   &  $0.12\pm 0.12$   &  $0.06^{+0.07}_{-0.06}$   \\
 Higgs  &  $2.7\pm 1.3$   &  $2.7\pm 1.5$   &  $1.5\pm 1.0$   &  $0.71\pm 0.29$   &  $0.57\pm 0.34$   &  $0.5\pm 0.5$   &  $0.6\pm 0.4$   &  $0.5\pm 0.5$   \\
 Reducible  &  $73^{+20}_{-17}$   &  $261 \pm 70$   &  $47^{+15}_{-13}$   &  $19\pm 5$   &  $71\pm 9$   &  $22.7\pm 2.8$   &  $630^{+9}_{-12}$   &  $162^{+6}_{-8}$   \\
\midrule
 Total SM  &  $191^{+24}_{-22}$   &  $794 \pm 86$   &  $69^{+15}_{-14}$   &  $98\pm 10$   &  $89^{+10}_{-9}$   &  $28.2\pm 3.2$   &  $648^{+10}_{-13}$   &  $166^{+6}_{-8}$   \\
 Data  &  $228$   &  $792$   &  $79$    &  $110$   &  $82$    &  $26$   &  $656$    &  $158$    \\
 \midrule
 CL$_b$     &  $0.90$   &  $0.49$   &  $0.72$  &  $0.79$   &  $0.30$   &  $0.37$   &  $0.61$  &  $0.30$ \\
\bottomrule 
   \end{tabular}
 }
\end{center}
\end{table}
\begin{figure}[t]
\centering
\includegraphics[width=0.50\textwidth]{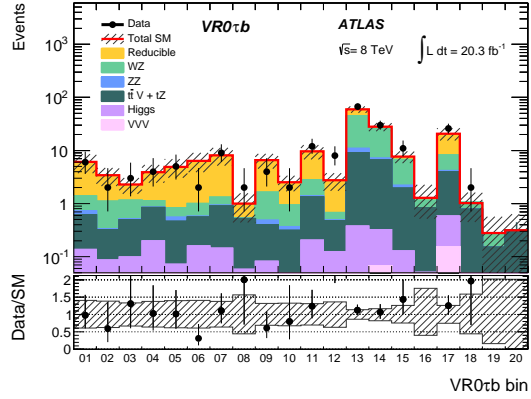}
\caption{Number of expected and observed events in the validation region VR0$\tau$b. 
Also shown are the respective contributions of the various background processes as described in the legend. 
The uncertainty band includes both the statistical and systematic uncertainties on the SM prediction. 
\label{fig:VR0b-summary}}
\end{figure}
\begin{figure}[!h]
\centering
\subfigure[]{\includegraphics[width=0.49\textwidth]{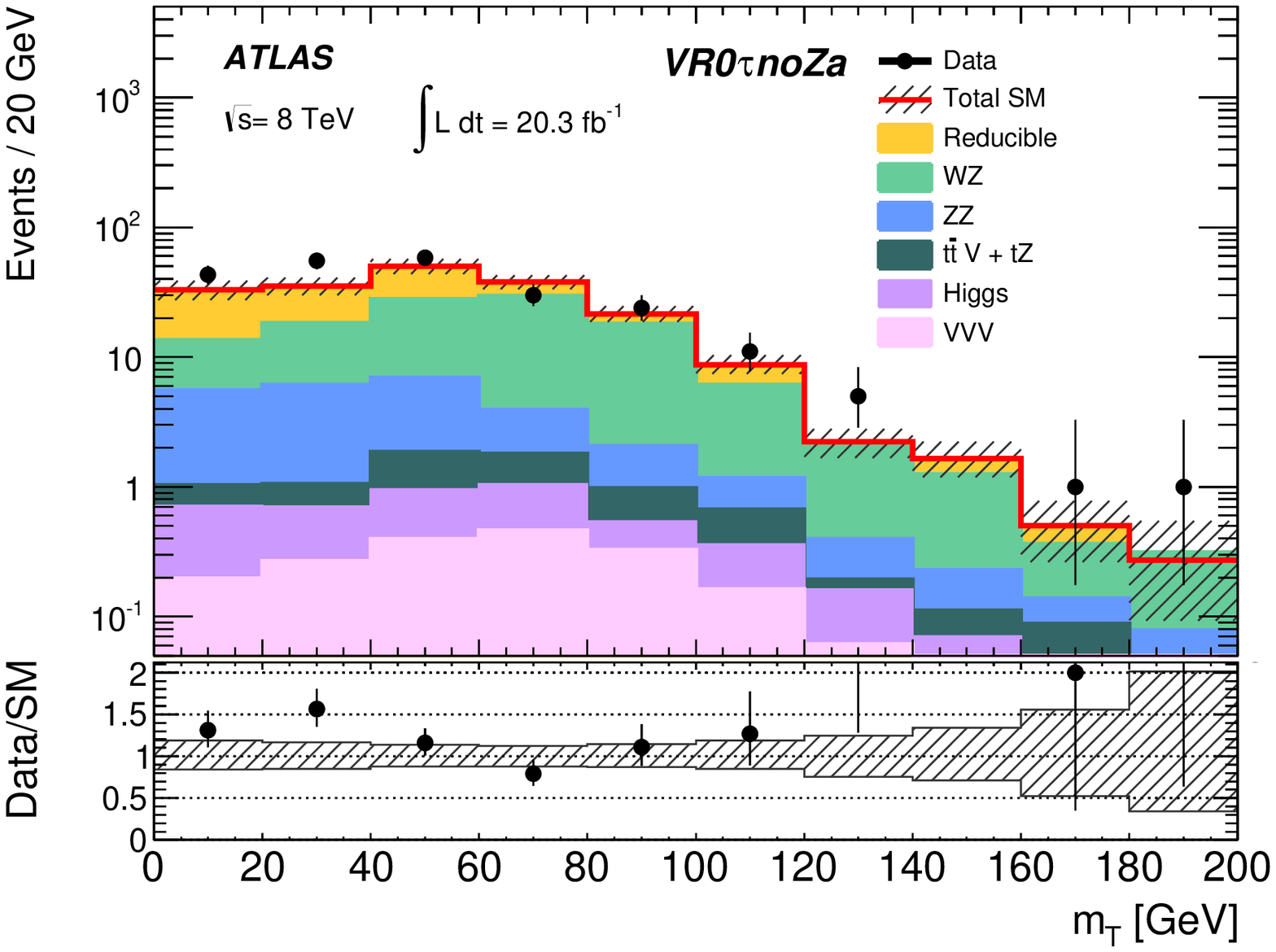}}
\subfigure[]{\includegraphics[width=0.49\textwidth]{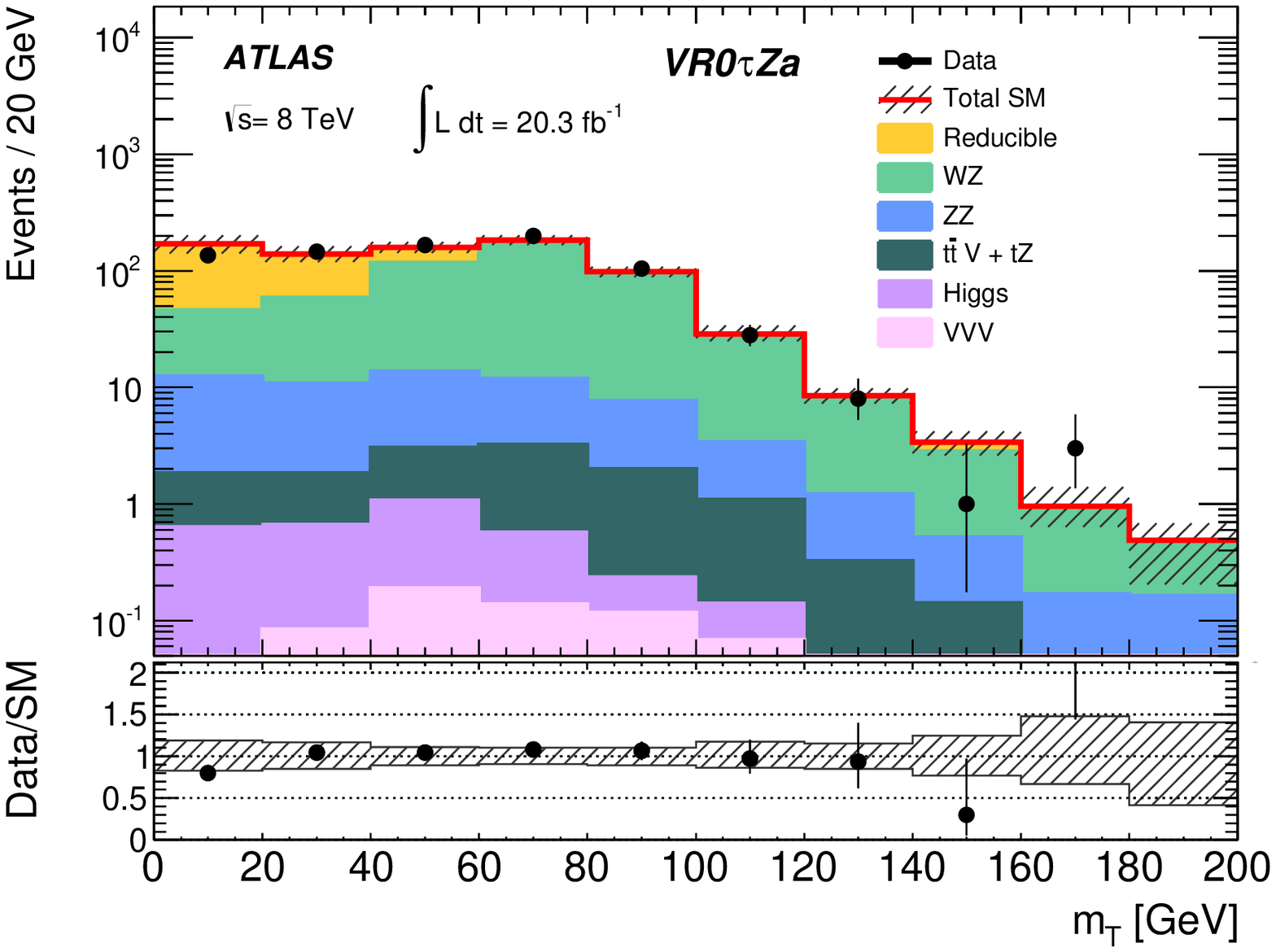}}
\subfigure[]{\includegraphics[width=0.49\textwidth]{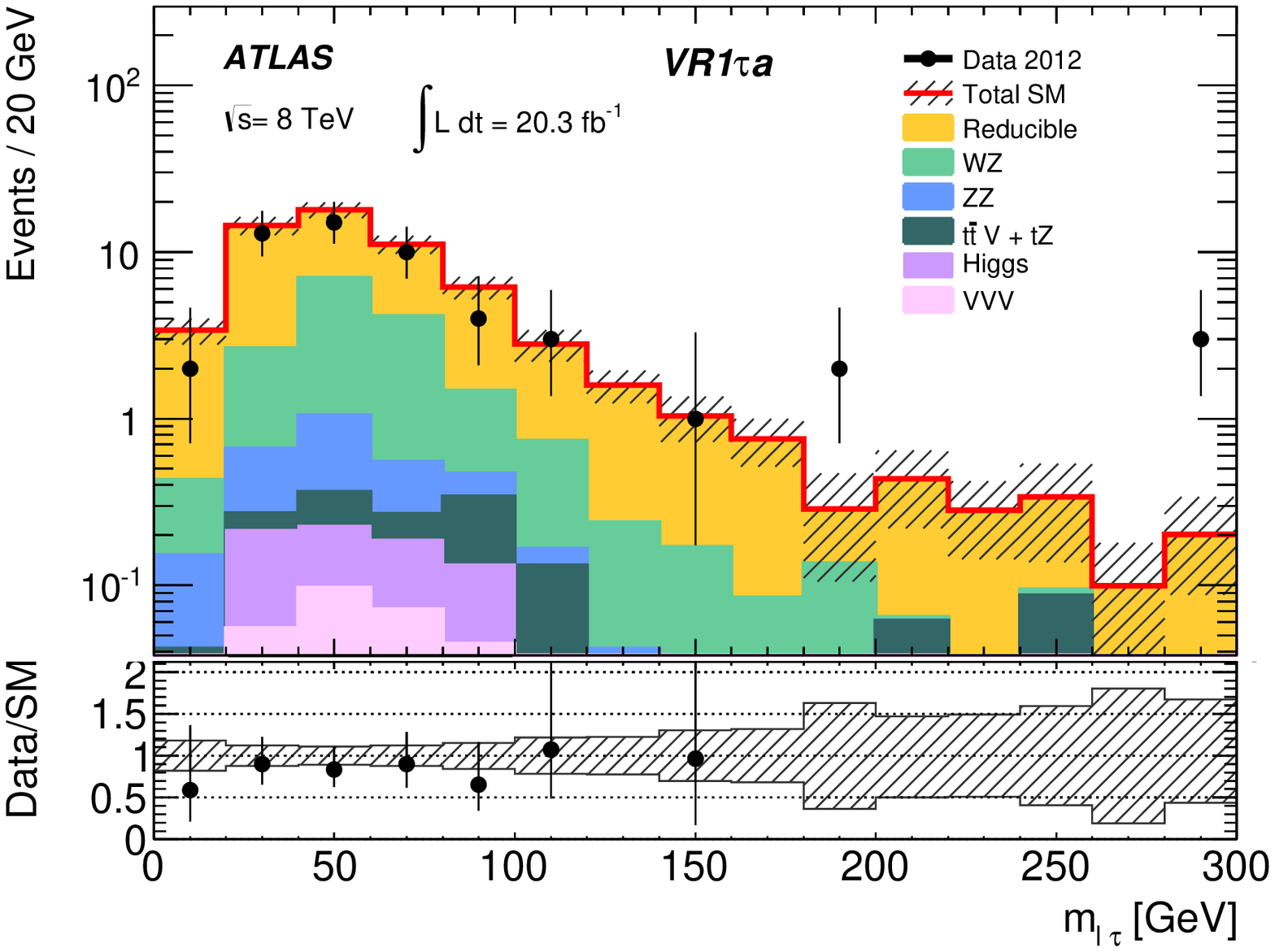}}
\subfigure[]{\includegraphics[width=0.49\textwidth]{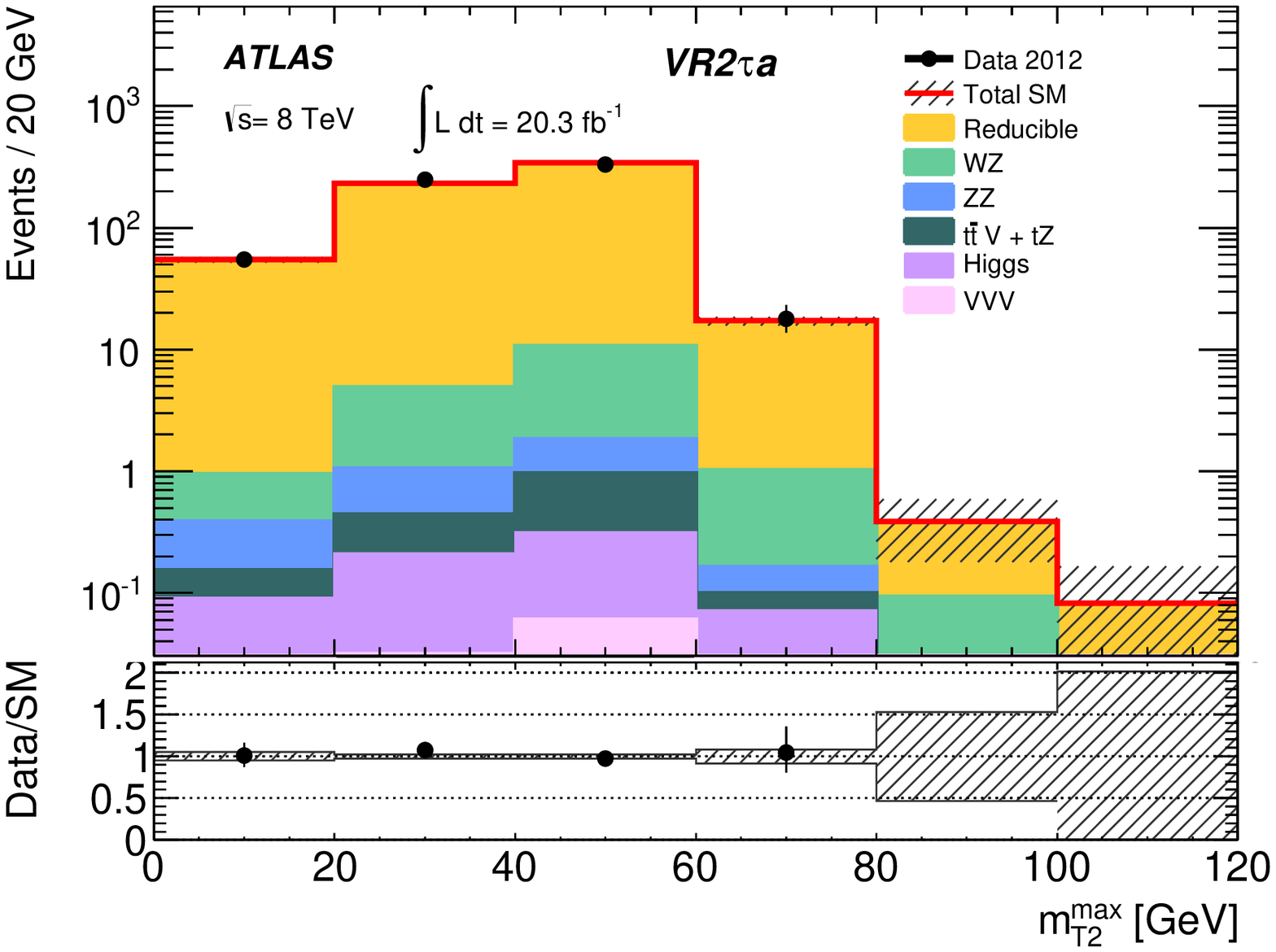}}
\caption{
For events in the low-$\met$ validation regions, the $\mt$ distribution in (a) VR0$\tau$noZa, (b) VR0$\tau$Za, (c) the $\mlt$ distribution in VR1$\tau$a and (d) the $\mttwo$ distribution in VR2$\tau$a, see text for details. 
Also shown are the respective contributions of the various background processes as described in the legend. 
The uncertainty band includes both the statistical and systematic uncertainties on the SM prediction. 
The last bin in each distribution includes the overflow.
\label{fig:VRplotsa}}
\end{figure}
\begin{figure}[h]
\centering
\subfigure[]{\includegraphics[width=0.49\textwidth]{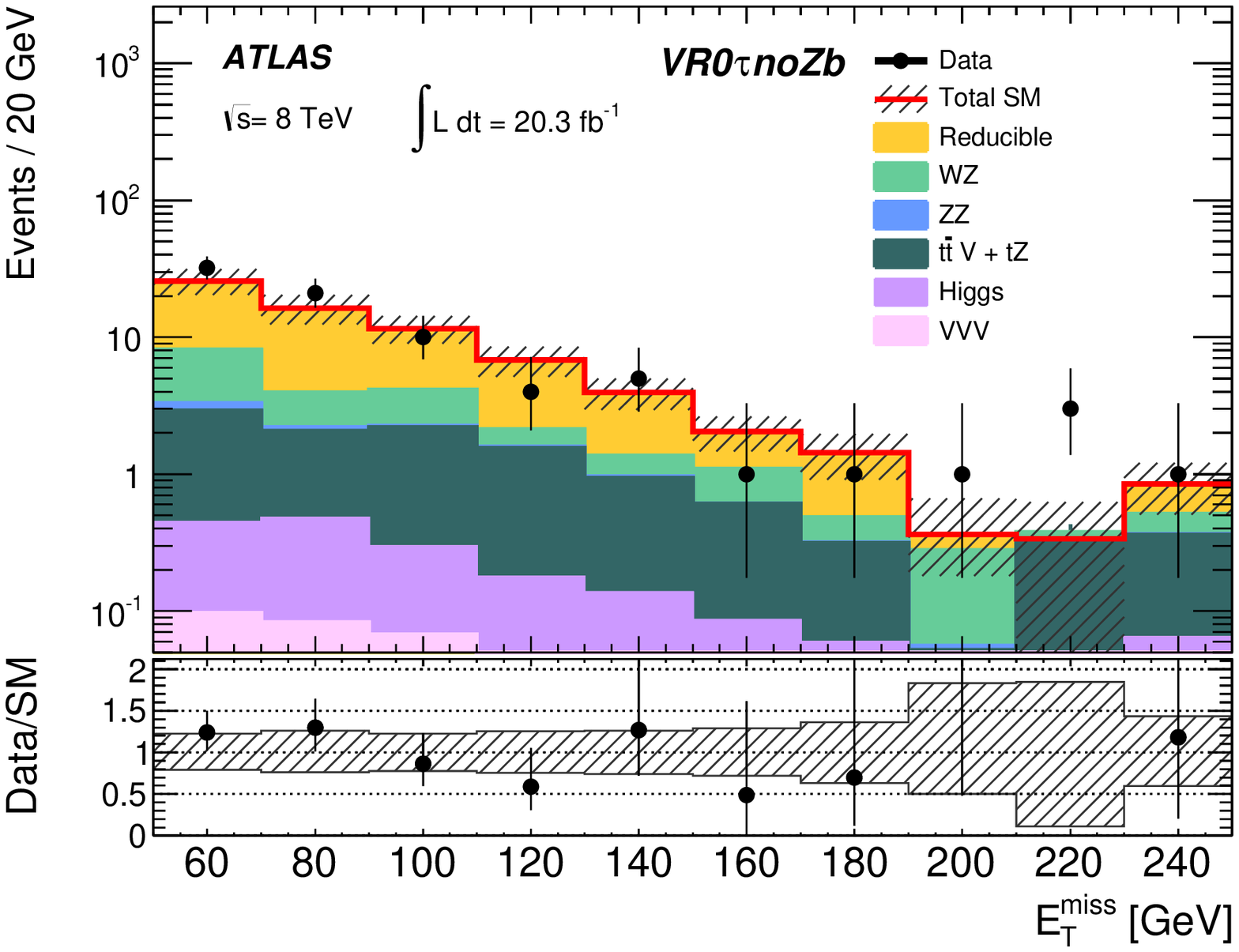}}
\subfigure[]{\includegraphics[width=0.49\textwidth]{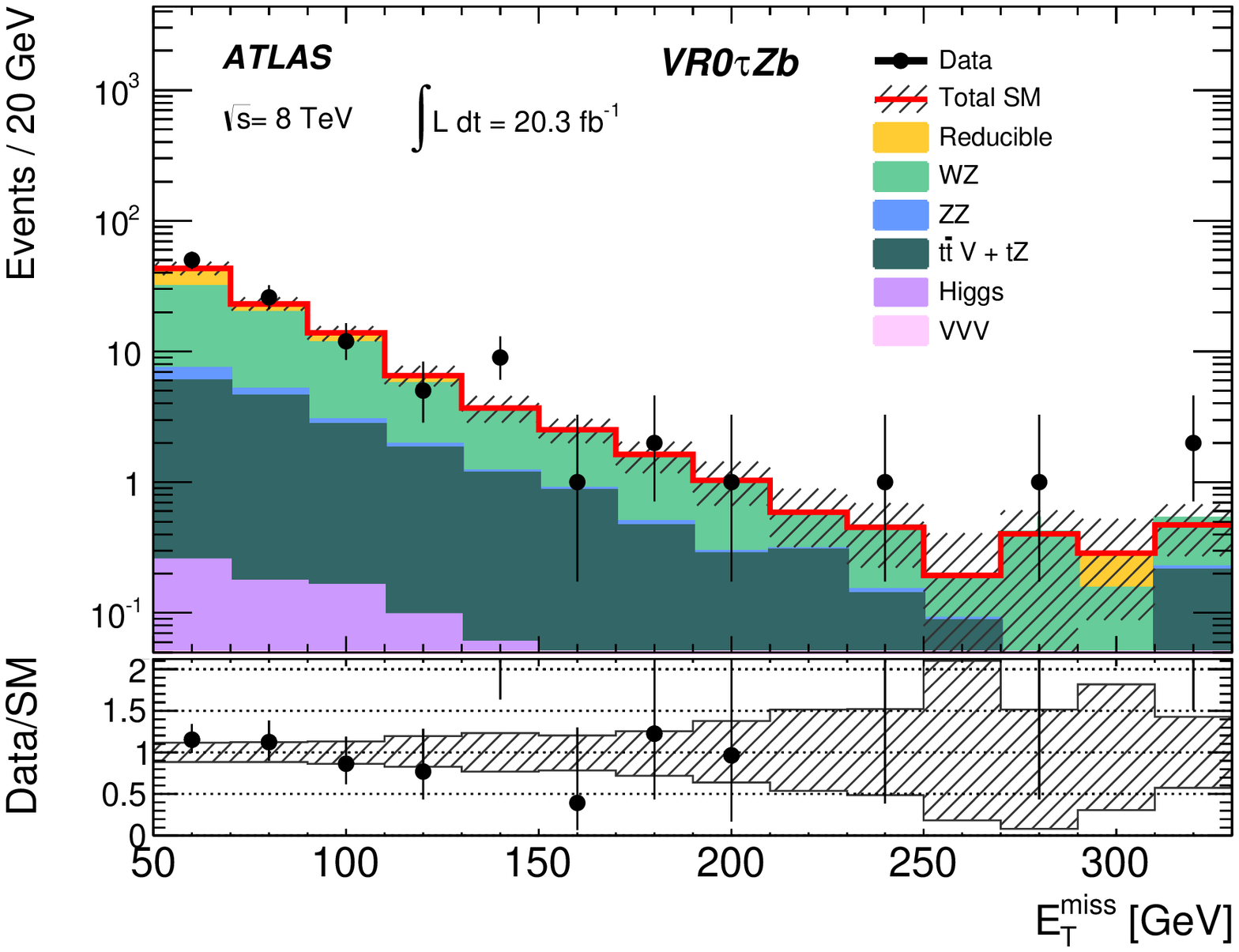}}
\subfigure[]{\includegraphics[width=0.49\textwidth]{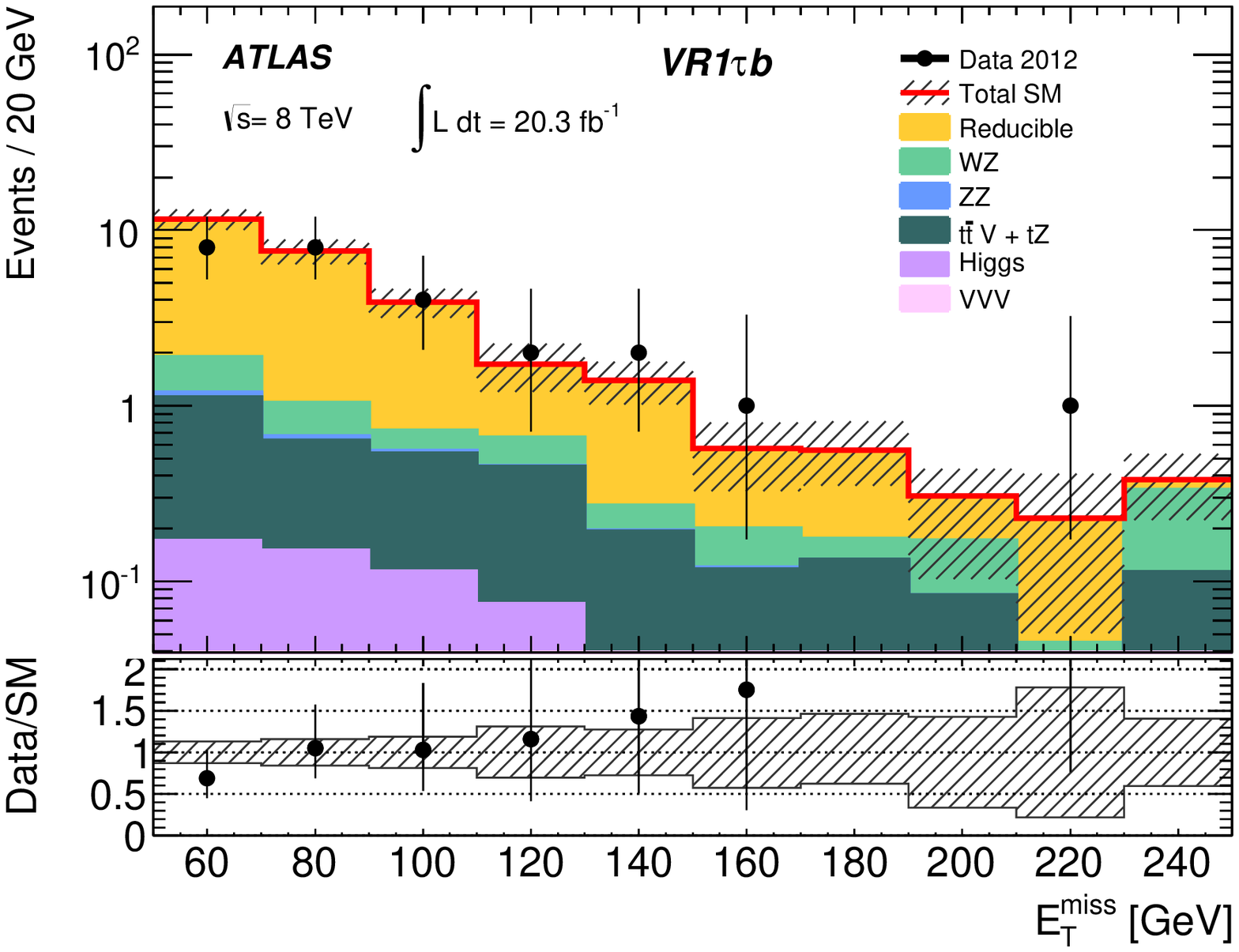}}
\subfigure[]{\includegraphics[width=0.49\textwidth]{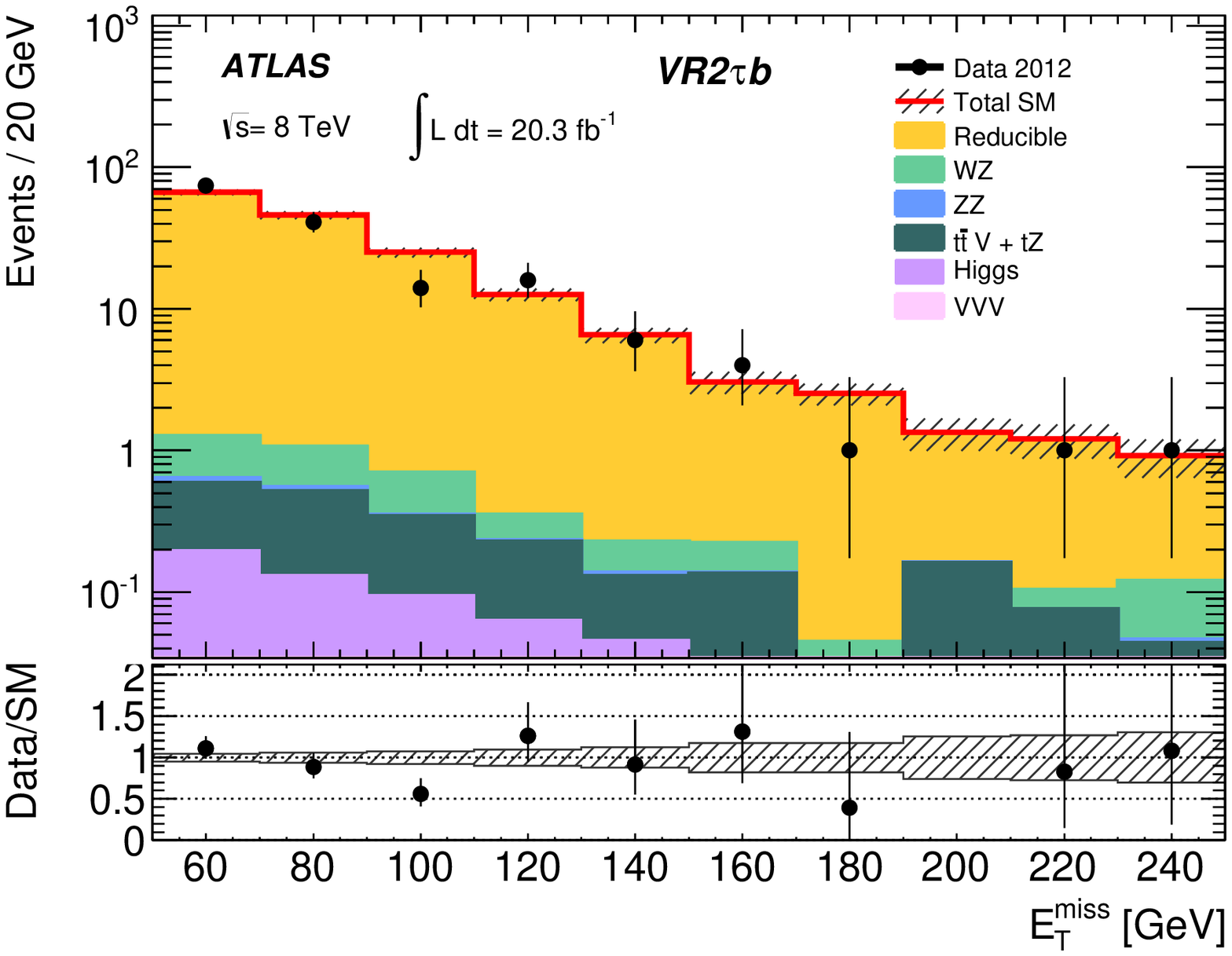}}
\caption{
For events in the high-$\met$ + $b$-tagged jet validation regions, the $\met$ distribution in (a) VR0$\tau$noZb, (b) VR0$\tau$Zb, (c) VR1$\tau$b and (d) VR2$\tau$b. 
Also shown are the respective contributions of the various background processes as described in the legend. 
The uncertainty band includes both the statistical and systematic uncertainties on the SM prediction. 
The last bin in each distribution includes the overflow.
\label{fig:VRplotsb}}
\end{figure}

In the validation regions, the observed data count and SM expectations are in good agreement within statistical and systematic uncertainties, as shown in table~\ref{tab:bkgLepVR} and figure~\ref{fig:VR0b-summary}. 
The CL$_b$ value~\cite{Read:2002hq}, using a  profile likelihood ratio as a test statistic~\cite{Cowan:2010js}, verifies  the compatibility of the observation with the background prediction. 
Values of CL$_b$ above (below) 0.5 indicate the observed level of agreement above (below) the expected yield. 
The $\mt$ distributions in VR0$\tau$noZa and VR0$\tau$Za along with the $\mlt$ distribution in VR1$\tau$a and $\mttwo$ distribution in VR2$\tau$a are shown in figure~\ref{fig:VRplotsa}, while the $\met$ distributions in the high-$\met$ + $b$-tagged jet validation regions are presented in figure~\ref{fig:VRplotsb}.

While the results from the validation regions are not used to derive correction factors for the background, potential signal contamination is assessed. 
It is found to be at the sub-percent level for most of the SUSY scenarios considered except for some characterised by low chargino mass. 
These scenarios would, however, lead to a detectable signal in the signal regions.

\section{Results and interpretations\label{sec:results}}

The observed number of events in each signal region is shown in tables~\ref{tab:SR} and \ref{tab:SR2} along with the background expectations and uncertainties.
The uncertainties include both the statistical and systematic components described in section~\ref{sec:systematics}.
A summary of the dominant systematic uncertainties in each signal region is given in table~\ref{tab:systSumm}.

\begin{table}[ht]
\centering
\caption{
Expected numbers of SM background events and observed numbers of data events in the signal regions SR0$\tau$a-bin01--bin12 for 20.3$\,$fb$^{-1}$. 
Statistical and systematic uncertainties are included as described in section~\ref{sec:systematics}. 
Also shown are the one-sided $p_0$-values and the upper limits at 95\% CL on the expected and observed number of beyond-the-SM events ($N^{95}_{\rm exp}$ and $N^{95}_{\rm obs}$) for each signal region, calculated using pseudo-experiments and the CL$_s$ prescription. 
For $p_0$-values below 0.5, the observed number of standard deviations, $\sigma$, is also shown in parentheses. 
 \label{tab:SR}}
 \scriptsize{
 \renewcommand\arraystretch{1.5}
 \begin{tabular}{ l   ccccccc  }
 \toprule
Sample   & SR0$\tau$a-bin01  & SR0$\tau$a-bin02  & SR0$\tau$a-bin03  & SR0$\tau$a-bin04  & SR0$\tau$a-bin05  & SR0$\tau$a-bin06  \\
\midrule
 $WZ$  &  $13.2^{+3.4}_{-3.2}$   &  $3.0\pm 1.4$   &  $7.8\pm 1.6$   &  $4.5^{+1.1}_{-1.0}$   &  $6.3\pm 1.6$   &  $3.7\pm 1.6$   \\
 $ZZ$  &  $1.4^{+0.6}_{-0.5}$   &  $0.12\pm 0.06$   &  $0.40\pm 0.14$   &  $0.20\pm 0.18$   &  $1.5\pm 0.5$   &  $0.25^{+0.14}_{-0.11}$   \\
 $\ttbar V$ + $tZ$   &  $0.14\pm 0.05$   &  $0.07\pm 0.04$   &  $0.04^{+0.05}_{-0.04}$   &  $0.14\pm 0.13$   &  $0.11\pm 0.08$   &  $0.047^{+0.022}_{-0.021}$   \\
 $VVV$  &  $0.33\pm 0.33$   &  $0.10\pm 0.10$   &  $0.19\pm 0.19$   &  $0.6\pm 0.6$   &  $0.26^{+0.27}_{-0.26}$   &  $0.24\pm 0.24$    \\
 Higgs  &  $0.66\pm 0.26$   &  $0.15\pm 0.08$   &  $0.64\pm 0.22$   &  $0.46^{+0.18}_{-0.17}$   &  $0.36^{+0.14}_{-0.15}$   &  $0.33^{+0.13}_{-0.12}$   \\
 Reducible  &  $6.7\pm 2.4$   &  $0.8\pm 0.4$   &  $1.6^{+0.7}_{-0.6}$   &  $2.7\pm 1.0$   &  $4.3^{+1.6}_{-1.4}$   &  $2.0\pm 0.8$    \\
\midrule
 Total SM   &  $23\pm 4$   &  $4.2\pm 1.5$   &  $10.6\pm 1.8$   &  $8.5^{+1.7}_{-1.6}$   &  $12.9^{+2.4}_{-2.3}$   &  $6.6^{+1.9}_{-1.8}$     \\
 Data  &  $ 36 $  &  $ 5 $  &  $ 9 $  &  $ 9 $  &  $ 11 $  &  $ 13 $  \\
 \midrule
 $p_0$ ($\sigma$)  &  0.02 (2.16)  &  0.35 (0.38)  &  0.50   &  0.40 (0.26)  &  0.50  &  0.03 (1.91)   \\
 $N_{\rm exp}^{95}$  &  $14.1^{+5.6}_{-3.6}$  &  $6.2^{+2.5}_{-1.7}$  &  $8.4^{+3.1}_{-2.3}$ &  $7.7^{+3.1}_{-2.1}$  &  $9.0^{+3.6}_{-2.5}$  &  $8.0^{+3.2}_{-1.9}$   \\
 $N_{\rm obs}^{95}$  &  26.8 & 6.9 & 7.3 & 8.4 & 7.9 & 14.4 \\
\bottomrule
\toprule
Sample     & SR0$\tau$a-bin07 & SR0$\tau$a-bin08  & SR0$\tau$a-bin09  & SR0$\tau$a-bin10  & SR0$\tau$a-bin11  & SR0$\tau$a-bin12  \\
\midrule
 $WZ$  &  $7.6\pm 1.3$   &  $0.30^{+0.25}_{-0.24}$   &  $16.2^{+3.2}_{-3.1}$   &  $13.1^{+2.5}_{-2.6}$   &  $19\pm 4$   &  $3.7\pm 1.2$    \\
 $ZZ$  &  $0.55^{+0.16}_{-0.14}$   &  $0.012^{+0.008}_{-0.007}$   &  $1.43^{+0.32}_{-0.28}$   &  $0.60^{+0.12}_{-0.13}$   &  $0.7\pm 1.2$   &  $0.14\pm 0.09$    \\
 $\ttbar V$ + $tZ$   &  $0.04^{+0.15}_{-0.04}$  &  $0.12^{+0.13}_{-0.12}$   &  $0.16^{+0.09}_{-0.12}$   &  $0.12\pm 0.10$   &  $0.41^{+0.24}_{-0.22}$   &  $0.12\pm 0.11$   \\
 $VVV$  &  $0.9\pm 0.9$  &  $0.13^{+0.14}_{-0.13}$   &  $0.23^{+0.24}_{-0.23}$   &  $0.4\pm 0.4$   &  $0.6\pm 0.6$   &  $0.6\pm 0.6$    \\
 Higgs  &  $0.98^{+0.29}_{-0.30}$   &  $0.13\pm 0.06$   &  $0.32\pm 0.11$   &  $0.22^{+0.10}_{-0.11}$   &  $0.28\pm 0.12$   &  $0.12\pm 0.06$    \\
 Reducible  &  $4.0^{+1.5}_{-1.4}$  &  $0.40^{+0.27}_{-0.26}$   &  $4.1^{+1.3}_{-1.2}$   &  $1.9^{+0.9}_{-0.8}$   &  $5.7^{+2.1}_{-1.9}$   &  $0.9^{+0.5}_{-0.4}$    \\
\midrule
 Total SM    &  $14.1\pm 2.2$ &  $1.1\pm 0.4$   &  $22.4^{+3.6}_{-3.4}$   &  $16.4\pm 2.8$   &  $27\pm 5$   &  $5.5^{+1.5}_{-1.4}$    \\
 Data  &  $ 15 $  &  $ 1 $  &  $ 28 $  &  $ 24 $  &  $ 29 $  &  $ 8 $  \\
\midrule
 $p_0$ ($\sigma$)  &  0.37 (0.33) &  0.50  &  0.13 (1.12) &  0.07 (1.50)  &  0.39 (0.28)  &  0.21 (0.82)  \\
 $N_{\rm exp}^{95}$  &  $9.6^{+3.9}_{-2.5}$ & $3.7^{+1.5}_{-0.9}$ & $12.7^{+4.9}_{-3.5}$ & $11.3^{+4.5}_{-3.1}$ & $13.8^{+5.4}_{-3.7}$ & $6.9^{+2.9}_{-1.7}$ \\
 $N_{\rm obs}^{95}$  &  10.8 & 3.7 & 18.0 & 18.3 & 15.3 & 9.2 \\
 \bottomrule
 \end{tabular}
 }
\end{table}

\begin{table}[ht]
\centering
\caption{
Expected numbers of SM background events and observed numbers of data events in the signal regions SR0$\tau$a-bin13--bin20, SR0$\tau$b, SR1$\tau$, SR2$\tau$a and SR2$\tau$b for 20.3$\,$fb$^{-1}$. 
Statistical and systematic uncertainties are included as described in section~\ref{sec:systematics}. 
Also shown are the one-sided $p_0$-values and the upper limits at 95\% CL on the expected and observed number of beyond-the-SM events ($N^{95}_{\rm exp}$ and $N^{95}_{\rm obs}$) for each signal region, calculated using pseudo-experiments and the CL$_s$ prescription. 
For $p_0$-values below 0.5, the observed number of standard deviations, $\sigma$, is also shown in parentheses. 
 \label{tab:SR2}}
 \scriptsize{
 \renewcommand\arraystretch{1.5}
 \begin{tabular}{ l   cccccccc  }
  \toprule 
 Sample   & SR0$\tau$a-bin13  & SR0$\tau$a-bin14  & SR0$\tau$a-bin15  & SR0$\tau$a-bin16  & SR0$\tau$a-bin17  & SR0$\tau$a-bin18  \\
\midrule
 $WZ$  &  $613 \pm 65$   &  $207^{+33}_{-32}$  &  $58^{+12}_{-13}$   &  $3.9^{+1.6}_{-1.4}$   &  $50^{+7}_{-6}$   &  $2.3\pm 1.3$    \\
 $ZZ$  &  $29\pm 4$   &  $5.5\pm 1.5$  &  $3.5^{+1.1}_{-1.0}$   &  $0.12^{+0.08}_{-0.07}$   &  $2.4^{+0.7}_{-0.6}$   &  $0.08\pm 0.04$   \\
 $\ttbar V$ + $tZ$   &  $2.9^{+0.7}_{-0.6}$   &  $2.0^{+0.7}_{-0.6}$   &  $0.67^{+0.29}_{-0.28}$   &  $0.08^{+0.10}_{-0.08}$   &  $0.8\pm 0.5$   &  $0.15^{+0.16}_{-0.15}$    \\
 $VVV$  &  $1.3\pm 1.3$   &  $0.8\pm 0.8$  &  $1.0\pm 1.0$   &  $0.33\pm 0.33$   &  $3.2\pm 3.2$   &  $0.5\pm 0.5$   \\
 Higgs  &  $2.2\pm 0.7$   &  $0.98\pm 0.20$  &  $0.31\pm 0.11$   &  $0.033\pm 0.018$   &  $0.95\pm 0.29$   &  $0.05\pm 0.04$    \\
 Reducible  &  $68^{+21}_{-19}$   &  $2.2^{+1.9}_{-2.0}$  &  $1.2\pm 0.6$   &  $0.14^{+0.25}_{-0.14}$   &  $11.3^{+3.5}_{-3.2}$   &  $0.27\pm 0.20$    \\
\midrule
 Total SM     &  $715 \pm 70$   &  $219\pm 33$  &  $65\pm 13$   &  $4.6^{+1.7}_{-1.5}$   &  $69^{+9}_{-8}$   &  $3.4\pm 1.4$   \\
Data  &  $ 714 $  &  $ 214 $  &  $ 63 $  &  $ 3 $  &  $ 60 $  &  $ 1 $  \\
\midrule
 $p_0$ ($\sigma$)  &  0.50  &  0.50  &  0.50  &  0.50  &  0.50  &  0.50  \\
 $N_{\rm exp}^{95}$  &  $133^{+46}_{-36}$ & $66^{+24}_{-18}$ & $28.6^{+10.1}_{-7.2}$ & $5.9^{+2.6}_{-1.5}$ & $21.4^{+8.2}_{-5.6}$ & $4.8^{+2.0}_{-1.1}$ \\
 $N_{\rm obs}^{95}$  &  133 & 65 & 27.6 & 5.2 & 18.8 & 3.7 \\
\bottomrule
\toprule
Sample   & SR0$\tau$a-bin19  & SR0$\tau$a-bin20  & SR0$\tau$b  & SR1$\tau$   & SR2$\tau$a  & SR2$\tau$b   \\
\midrule
 $WZ$  &  $0.9\pm 0.4$   &  $0.12\pm 0.11$  &  $0.68\pm 0.20$   &  $4.6\pm 0.6$   &  $1.51^{+0.35}_{-0.33}$   &  $2.09^{+0.30}_{-0.31}$   \\
 $ZZ$  &  $0.021\pm 0.019$   &  $0.009\pm 0.009$   &  $0.028\pm 0.009$   &  $0.36\pm 0.08$   &  $0.049^{+0.016}_{-0.014}$   &  $0.135\pm 0.025$   \\
 $\ttbar V$ + $tZ$   &  $0.0023^{+0.0032}_{-0.0019}$   &  $0.012^{+0.016}_{-0.012}$  &  $0.17^{+0.32}_{-0.17}$   &  $0.16^{+0.18}_{-0.16}$   &  $0.21^{+0.27}_{-0.21}$   &  $0.023^{+0.015}_{-0.018}$   \\
 $VVV$  &  $0.08\pm 0.08$   &  $0.07^{+0.08}_{-0.07}$   &  $1.0\pm 1.0$   &  $0.5\pm 0.5$   &  $0.09\pm 0.09$   &  $0.031\pm 0.033$   \\
 Higgs  &  $0.007\pm 0.006$   &  $0.0009\pm 0.0004$  &  $0.49\pm 0.17$   &  $0.28\pm 0.12$   &  $0.021\pm 0.010$   &  $0.08\pm 0.04$   \\
 Reducible  &  $0.17^{+0.16}_{-0.15}$   &  $0.08^{+0.11}_{-0.08}$  &  $1.5\pm 0.4$   &  $4.3\pm 0.8$   &  $5.1\pm 0.7$   &  $4.9\pm 0.7$   \\
\midrule
 Total SM     &  $1.2\pm 0.4$   &  $0.29^{+0.18}_{-0.17}$   &  $3.8\pm 1.2$   &  $10.3\pm 1.2$   &  $6.9\pm 0.8$   &  $7.2^{+0.7}_{-0.8}$  \\
 Data  &  $ 0 $  &  $ 0 $  &  $ 3 $  &  $ 13 $  &  $ 6 $  &  $ 5 $  \\
 \midrule
 $p_0$ ($\sigma$)  &  0.50 &  0.50  &  0.50  &  0.19 (0.86)  &  0.50  &  0.50  \\
 $N_{\rm exp}^{95}$  &  $3.7^{+1.4}_{-0.7}$ & $3.0^{+0.8}_{-0.0}$ & $5.6^{+2.2}_{-1.4}$ & $8.1^{+3.2}_{-2.2}$ & $6.8^{+2.7}_{-1.9}$ & $6.7^{+2.8}_{-1.8}$ \\
 $N_{\rm obs}^{95}$  &  3.0 & 3.0 & 5.4 & 10.9 & 6.0 & 5.2 \\
\bottomrule
 \end{tabular} 
 }
\end{table}

\begin{table}[ht]
\centering
\caption{Summary of the dominant systematic uncertainties in the background estimates for each signal region.
Uncertainties are quoted relative to the total expected background. 
For the 20 bins of the SR0$\tau$a the range of the uncertainties is provided.
\label{tab:systSumm} }
\footnotesize{
    \begin{tabular}{ l c c c c c c }

    \toprule
                &  SR0$\tau$a & SR0$\tau$b  &    SR1$\tau$  & SR2$\tau$a           & SR2$\tau$b \\
    \midrule
Cross-section                          & 4--25\%    & 37\%      & 9\%         & 3.1\%      & 3.0\% \\
Generator                              & 3.2--35\%  & 11\%      & 3.1\%       & 6\%        & $<1$\% \\
Statistics on irreducible background   & 0.8--26\%  & 8\%       & 5\%         & 5\%        & 3.1\% \\
Statistics on reducible background     & 0.4--29\%  & 14\%      & 8\%         & 13\%       & 12\% \\
Electron misidentification probability & 0.3--10\%  & 1.3\%     & $<1$\%      &--        &--\\
Muon misidentification probability     & 0.1--24\%  & 2.2\%     & $<1$\%      &--        &--         \\
$\tau$ misidentification probability      &--           &--       & 8\%         & 4\%        & 5\% \\
    \bottomrule
    \end{tabular}
}
\end{table}

\begin{figure}[h]
\centering
\subfigure[]{\includegraphics[width=0.49\textwidth]{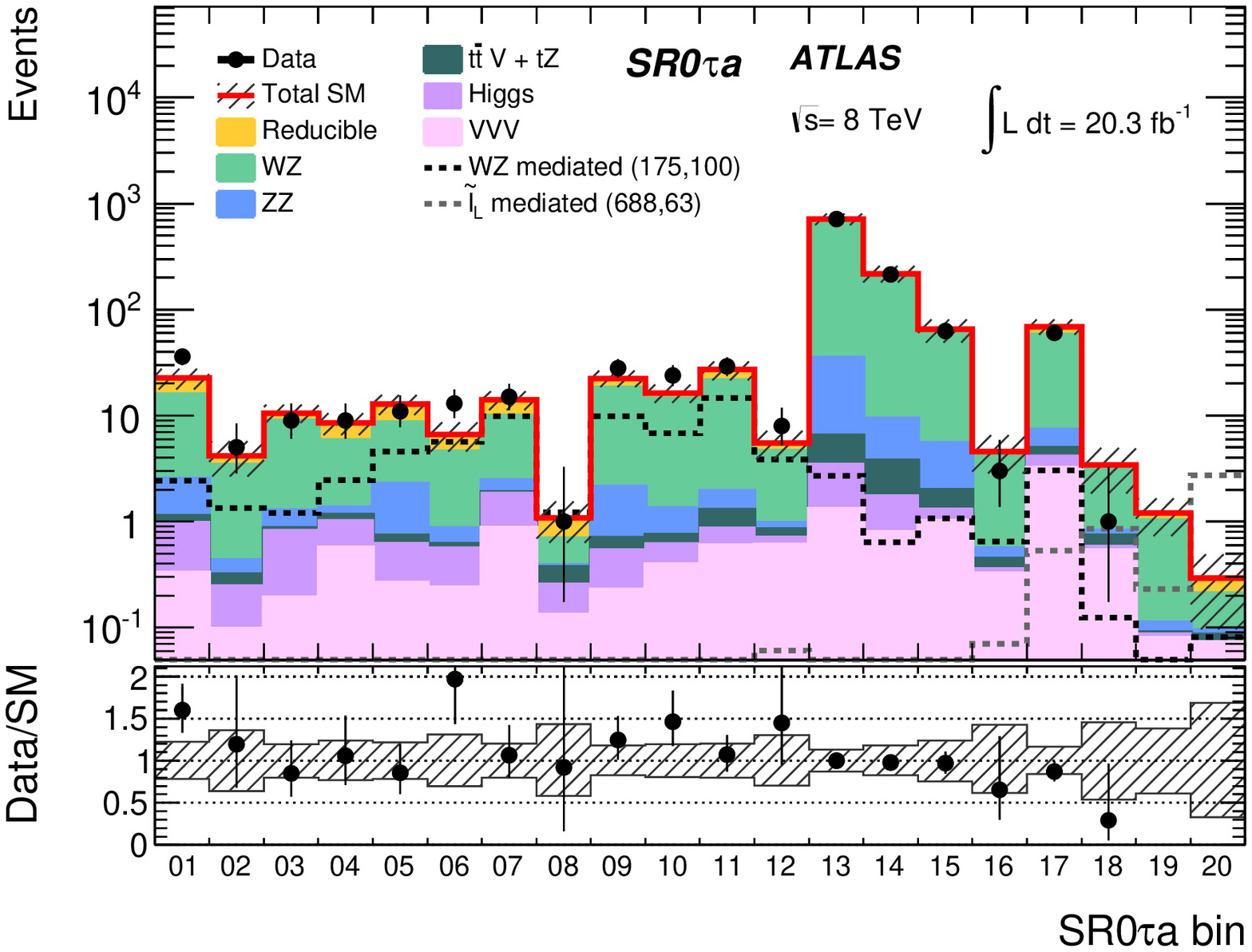}}
\subfigure[]{\includegraphics[width=0.49\textwidth]{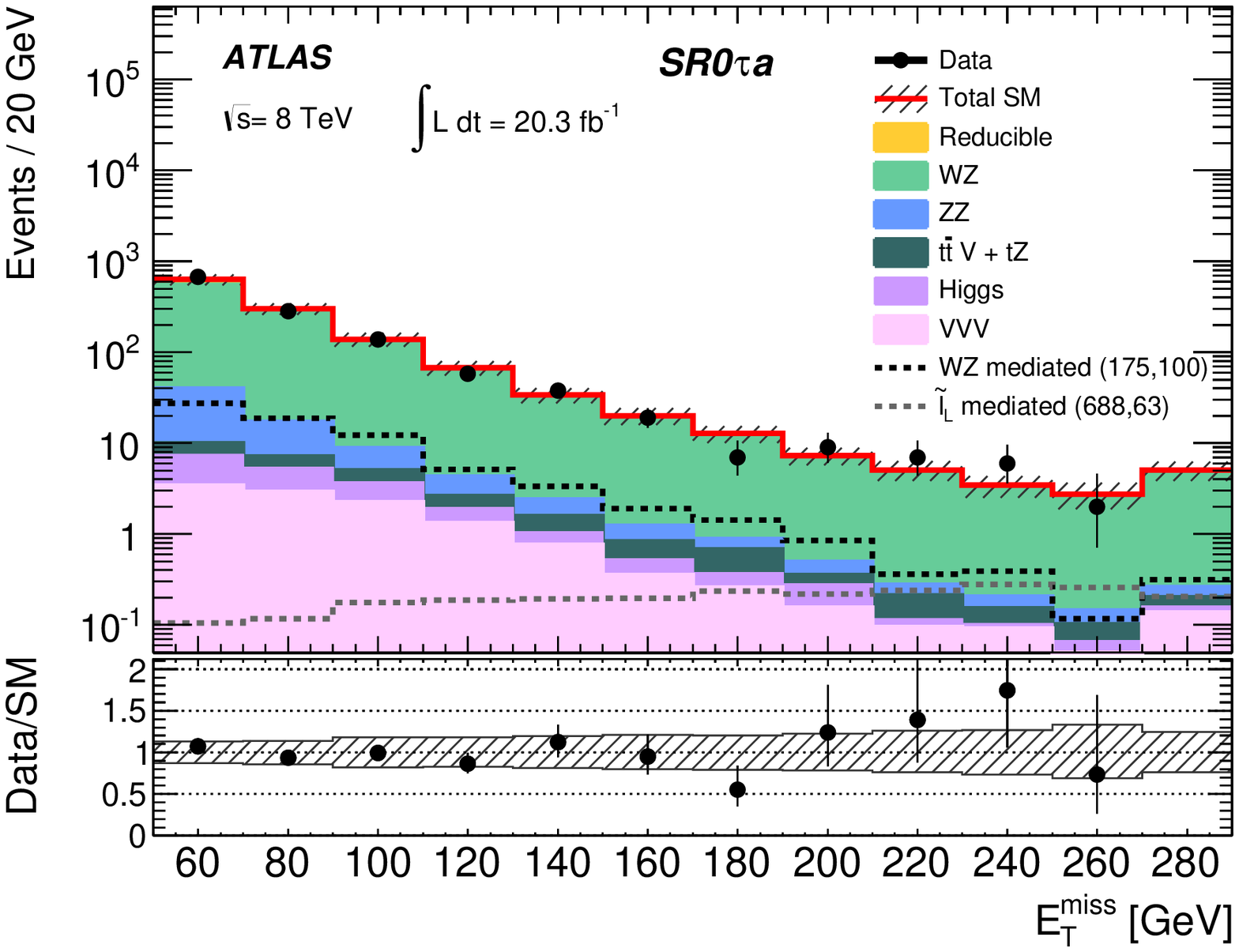}}
\subfigure[]{\includegraphics[width=0.49\textwidth]{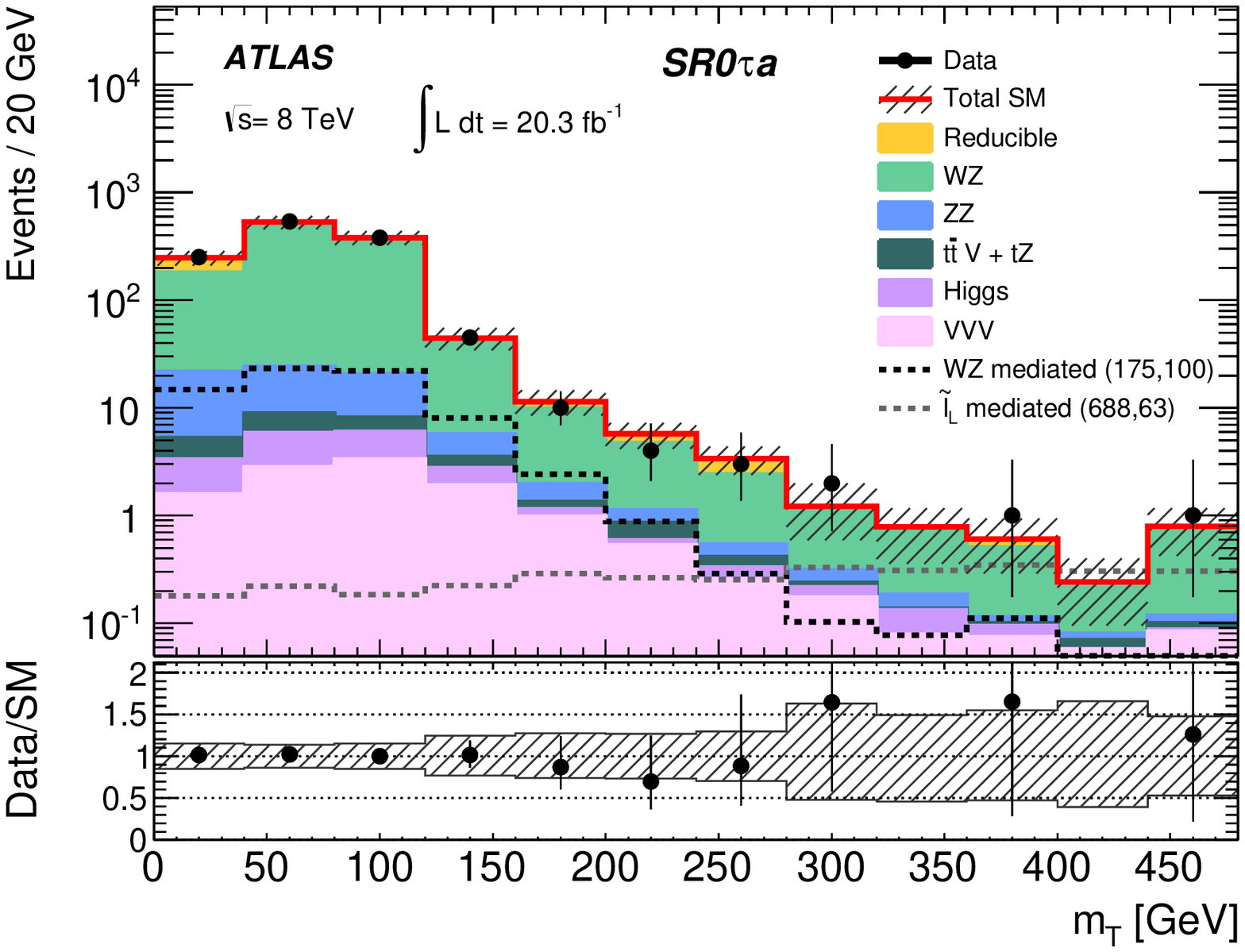}}
\subfigure[]{\includegraphics[width=0.49\textwidth]{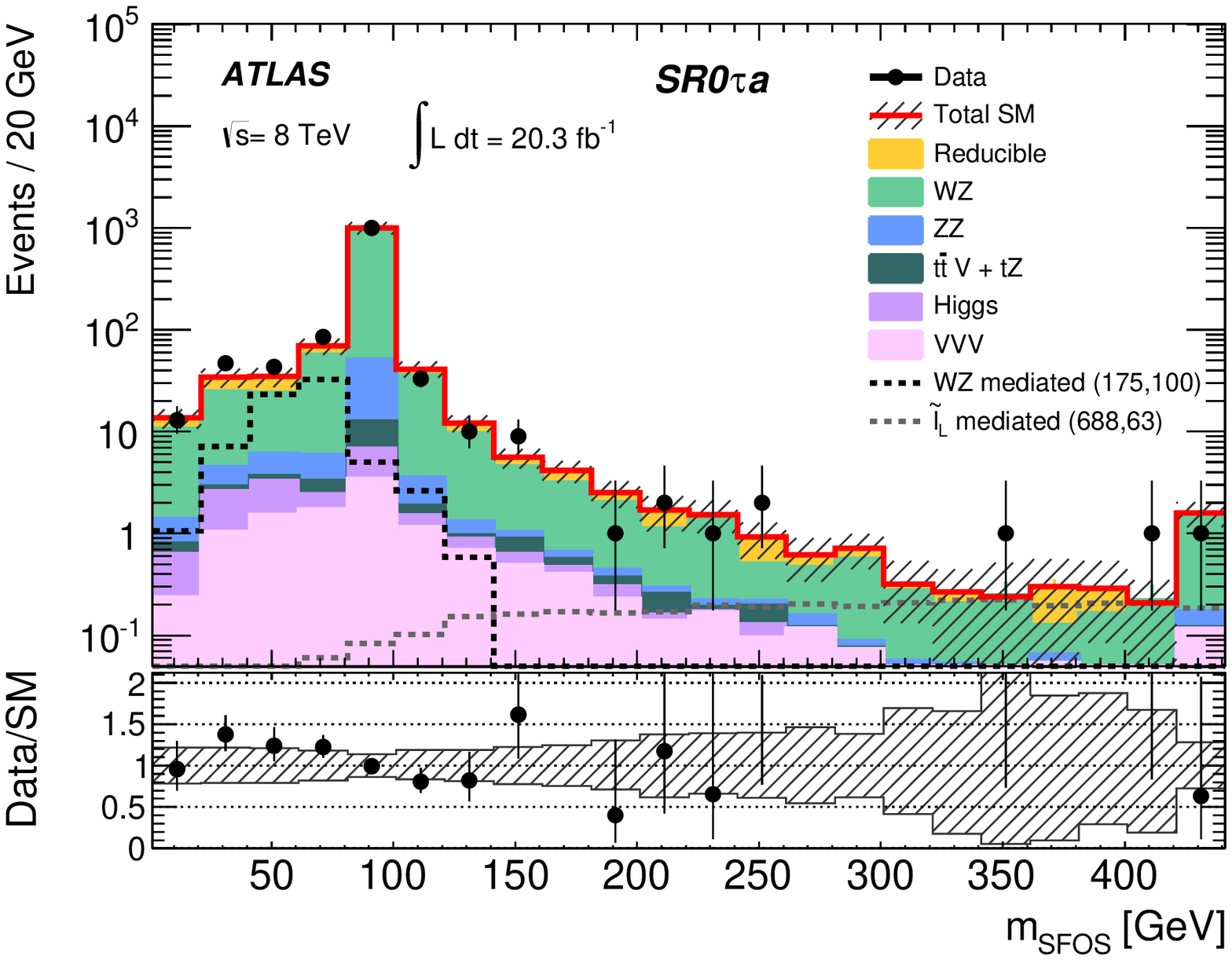}}
\caption{
Expected distributions of SM background events and observed data distributions in (a) the binned signal regions SR0$\tau$a.
The distributions of (b) \met, (c) \mt\ and (d) \msfos\ are shown in the summation of all SR0$\tau$a regions prior to the requirements on these variables. 
Also shown are the respective contributions of the various background processes as described in the legend. 
Both the statistical and systematic uncertainties are shown. 
The last bin in each distribution includes the overflow.
For illustration, the distributions of signal hypotheses are also shown.
\label{fig:SR0a}}
\end{figure}

Figure~\ref{fig:SR0a} shows the SM expectations and the observations in data in the individual SR0$\tau$a bins as well as the distribution of \met, \mt\ and \msfos\ in the combination of all SR0$\tau$a regions.
For illustration, the distributions are also shown for the $WZ$-mediated and $\slepL$-mediated simplified models.

\begin{figure}[h]
\centering
\subfigure[\label{fig:SR0b}]{\includegraphics[width=0.49\textwidth]{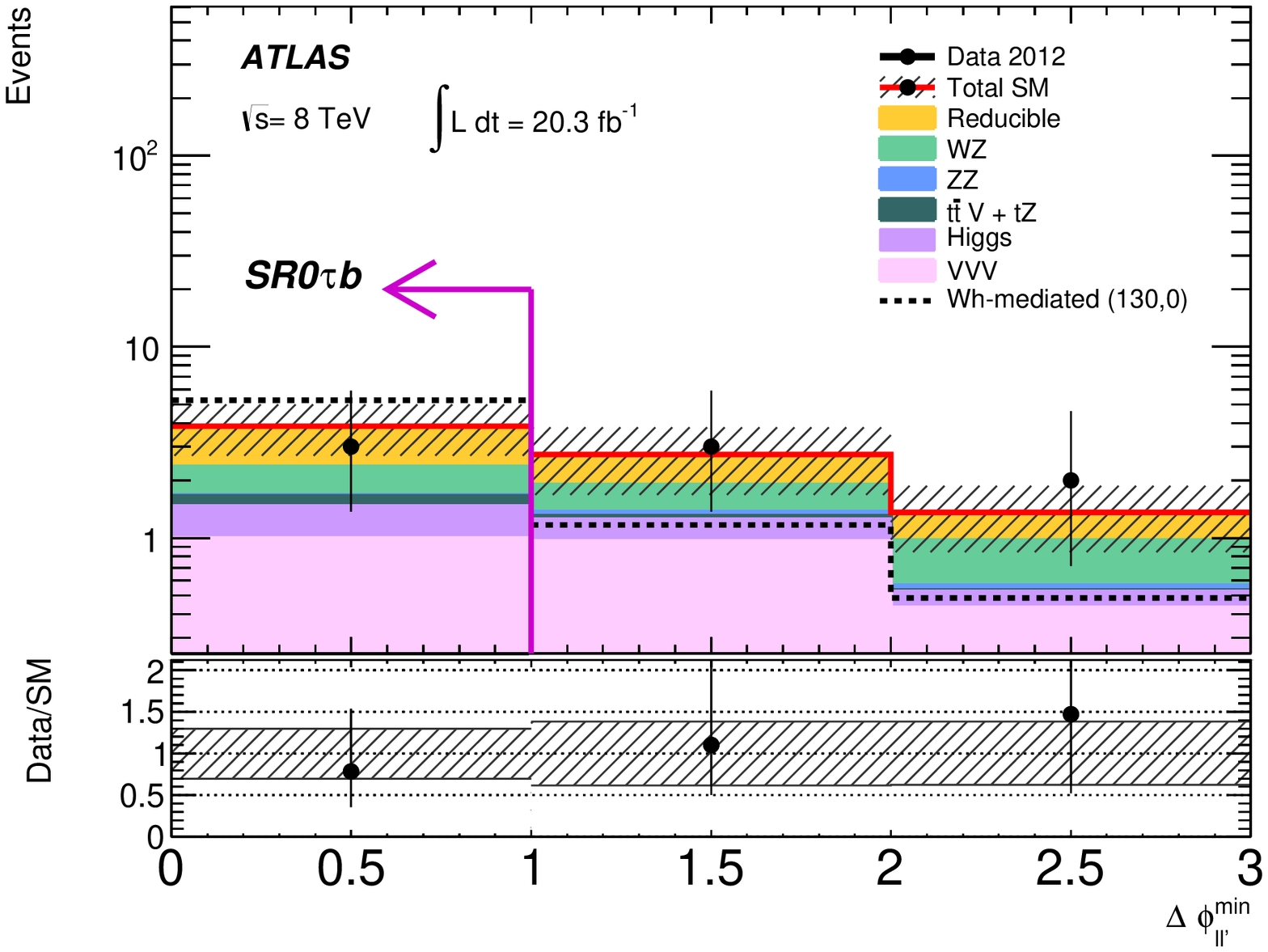}}
\subfigure[\label{fig:SR1}]{\includegraphics[width=0.49\textwidth]{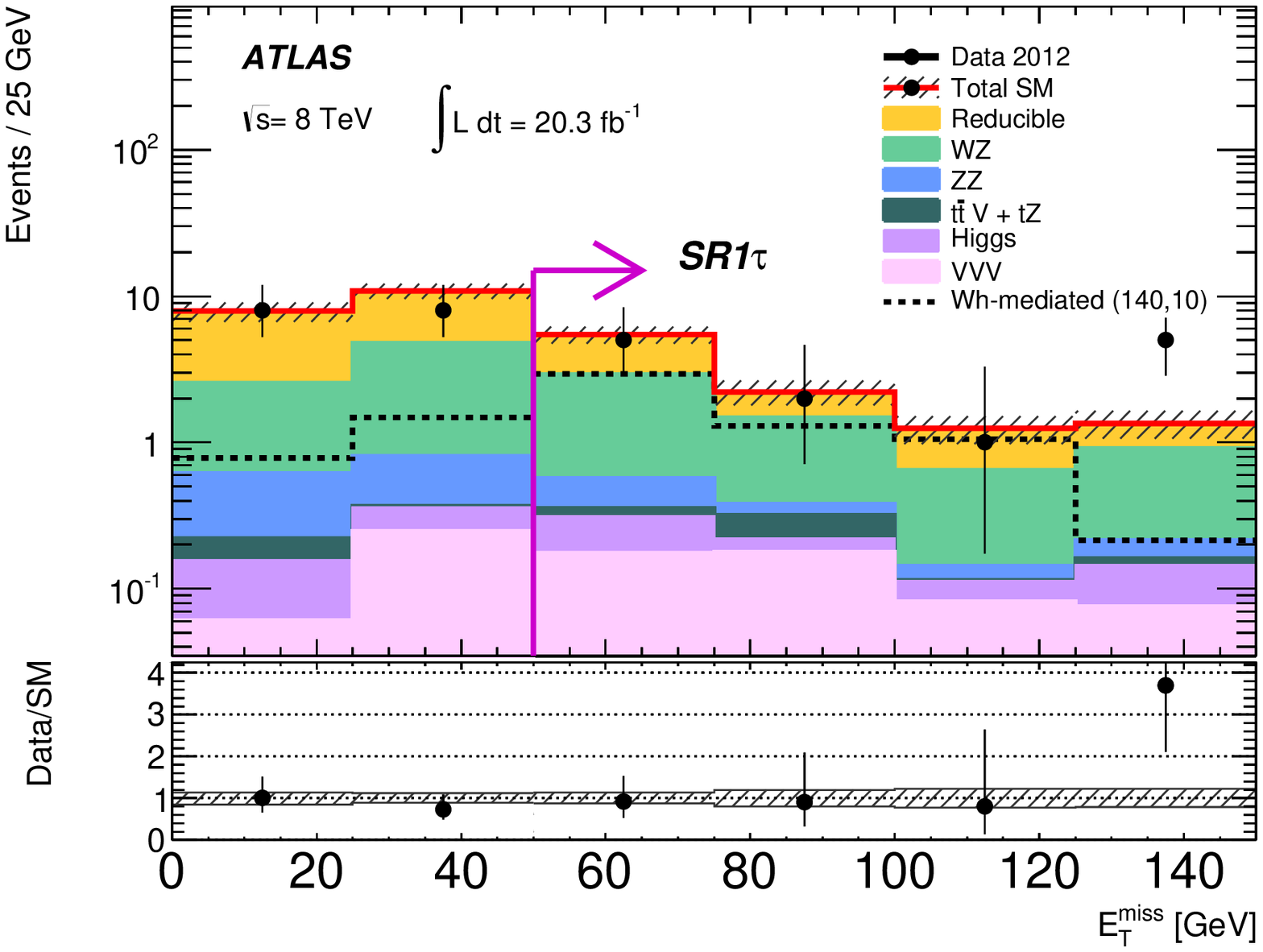}}
\subfigure[\label{fig:SR2a}]{\includegraphics[width=0.49\textwidth]{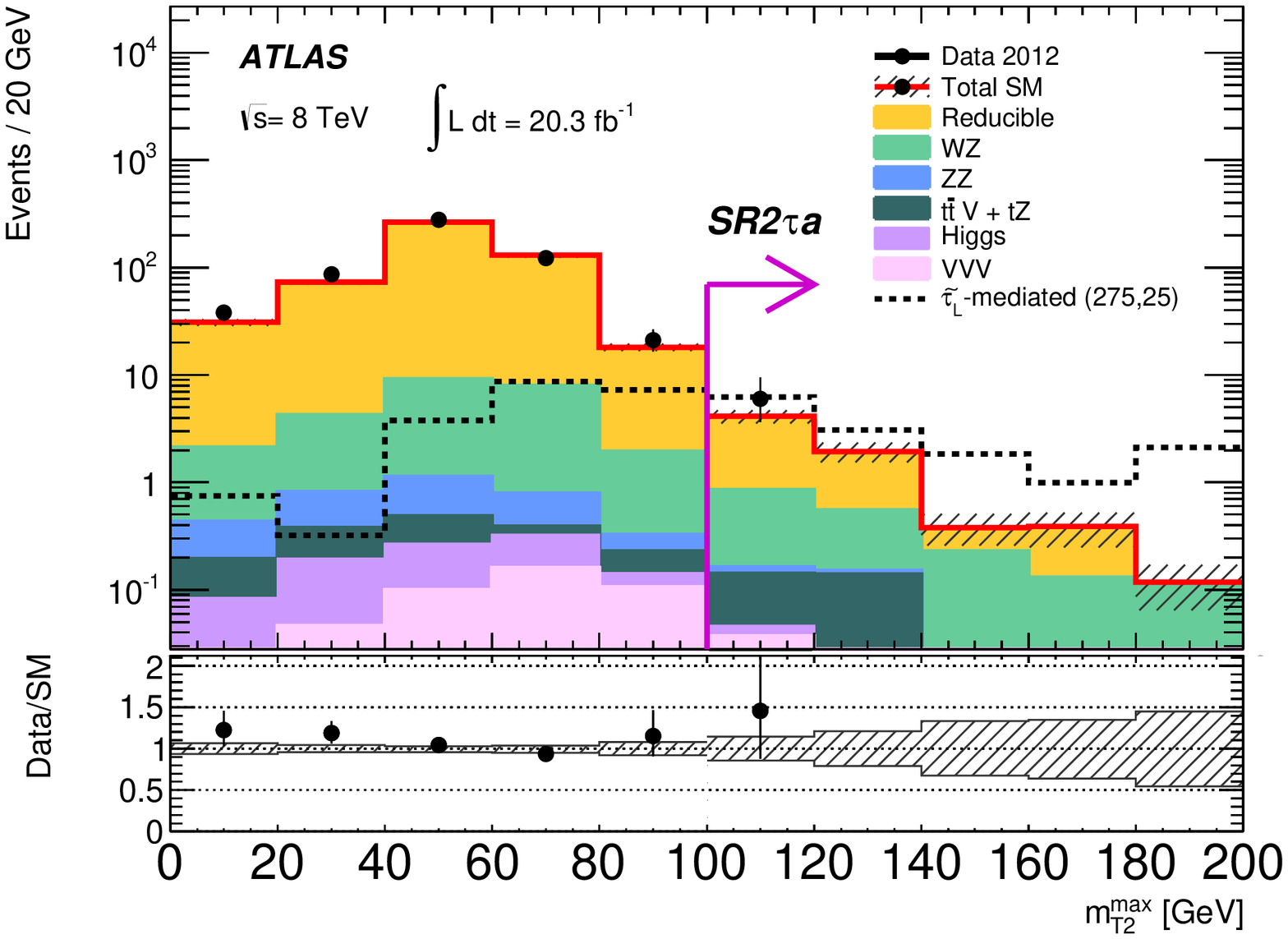}}
\subfigure[\label{fig:SR2b}]{\includegraphics[width=0.49\textwidth]{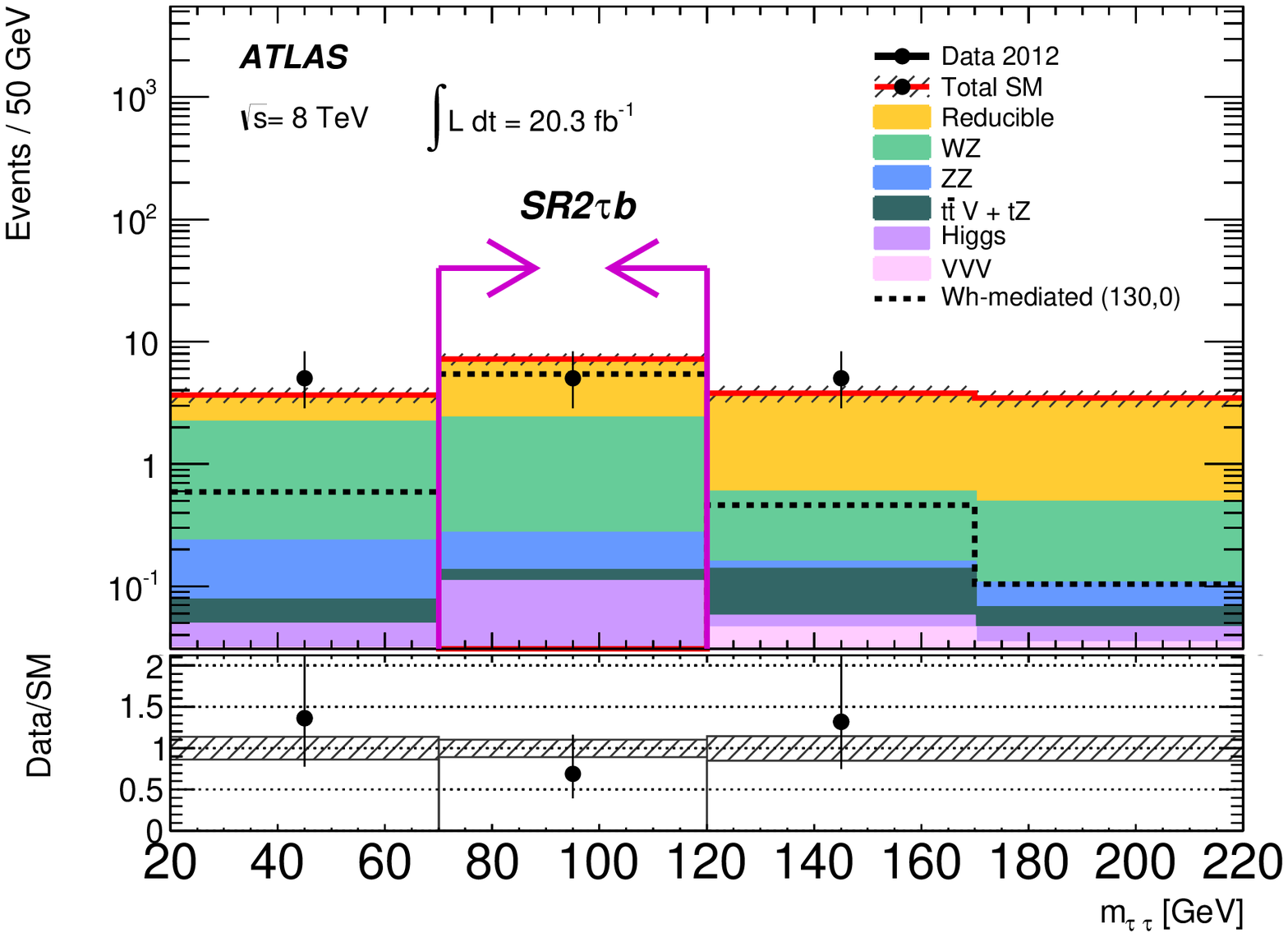}}
\caption{
Expected distributions of SM background events and observed data distributions for (a) \mindPhi,  (b) \met, (c) \mttwo\ and (d) \mtt\ variables in the SR0$\tau$b, SR1$\tau$, SR2$\tau$a and SR2$\tau$b regions respectively, prior to the requirements on these variables.
Arrows indicate the limits on the values of the variables used to define the signal regions.
Also shown are the respective contributions of the various background processes as described in the legend. 
Both the statistical and systematic uncertainties are shown.  
The last bin in each distribution includes the overflow.
The plots also show the distribution for signal hypotheses, where the parentheses following the simplified model denote the mass parameters in $\GeV$ as $(m(\chinoonepm,\ninotwo)$, $m(\ninoone)$).
\label{fig:SR012}}
\end{figure}

Figure~\ref{fig:SR012} shows the distributions of the quantities \mindPhi, \met, \mttwo\ and \mtt\ in the SR0$\tau$b, SR1$\tau$, SR2$\tau$a and SR2$\tau$b regions respectively, prior to the requirements on these variables. 
Also shown are the distributions of these quantities for signal hypotheses from the $\stauL$-mediated and $Wh$-mediated simplified models.

The number of observed events is consistent with the SM expectation in all signal regions, within uncertainties.
The one-sided $p_0$-value is calculated to quantify the probability of the SM background alone to fluctuate to the observed number of events or higher (shown in table~\ref{tab:SR} and table~\ref{tab:SR2}), and is truncated to 0.5 for $p_0\,$$>\,$0.5. 
Upper limits at 95\% CL on the expected and observed number of beyond the SM events ($N^{95}_{\rm exp}$ and $N^{95}_{\rm obs}$) for each signal region are calculated using the CL$_s$ prescription~\cite{Read:2002hq} and shown in table~\ref{tab:SR} and table~\ref{tab:SR2}. 
The profile likelihood ratio is used as a test statistic~\cite{Cowan:2010js} and sources of systematic uncertainties are treated as nuisance parameters. 
The $p_0$ and CL$_s$ values are calculated using pseudo-experiments.

The results obtained are used to derive limits on the simplified and pMSSM models described in section~\ref{sec:SUSY}.
Exclusion limits are calculated by statistically combining results from a number of disjoint signal regions.
For the $\slepL$-mediated, $WZ$-mediated and $\stauL$-mediated simplified models and for the pMSSM scenarios, SR0$\tau$a, SR0$\tau$b, SR1$\tau$ and SR2$\tau$a are statistically combined. 
For the $Wh$-mediated simplified model, the statistical combination of SR0$\tau$a, SR0$\tau$b, SR1$\tau$ and SR2$\tau$b is used. 
All experimental uncertainties are treated as correlated between regions and processes, with the exception of the experimental uncertainties on the reducible background, which are correlated between regions only. 
Theoretical uncertainties on the irreducible background and signal are treated as correlated between regions, while statistical uncertainties are treated as uncorrelated between regions and processes.  
The total systematic uncertainty on all SUSY signal processes is in the 10--20\% range, 
where $\sim\,$7\% originates from the uncertainty on the signal cross-section. 
The uncertainty due to changes in signal acceptance from varying the PDFs and the amount of initial-state radiation is found to be negligible compared to the total systematic uncertainty for the signal scenarios under consideration.
For the exclusion limits, the observed and expected 95\% CL limits are calculated using pseudo-experiments for each SUSY model point, taking into account the theoretical and experimental uncertainties on the SM background and the experimental uncertainties on the signal. 
The impact of the theoretical uncertainties on the signal cross-section is shown for the observed limit and where quoted, limits refer to the $-1\sigma$ variation on the observed limit.

\begin{figure}[h]
\centering
\subfigure[$\slepL$-mediated simplified model\label{fig:exclLimitsSimpa}]{\includegraphics[width=0.49\textwidth]{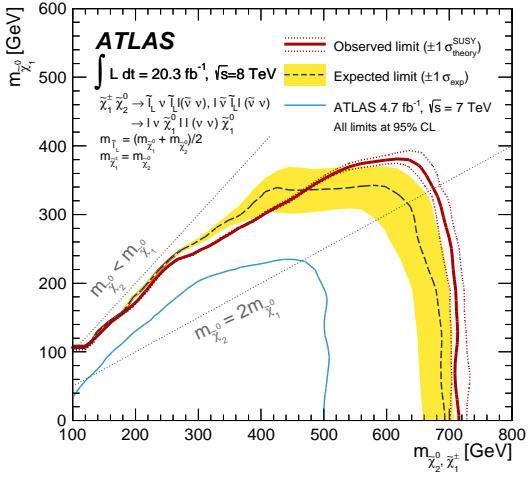}}
\subfigure[$WZ$-mediated simplified model\label{fig:exclLimitsSimpb}]{\includegraphics[width=0.49\textwidth]{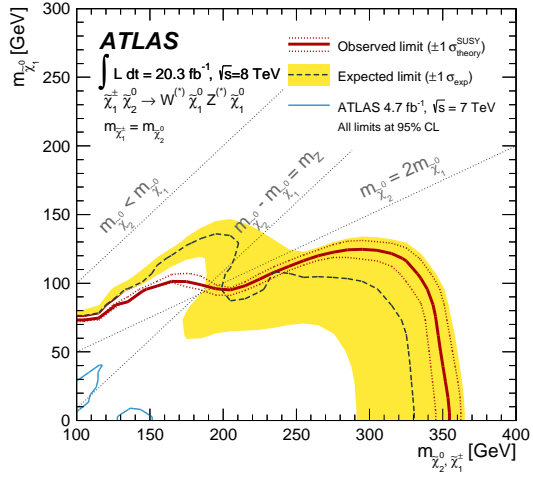}}
\subfigure[$\stauL$-mediated simplified model\label{fig:exclLimitsSimpc}]{\includegraphics[width=0.49\textwidth]{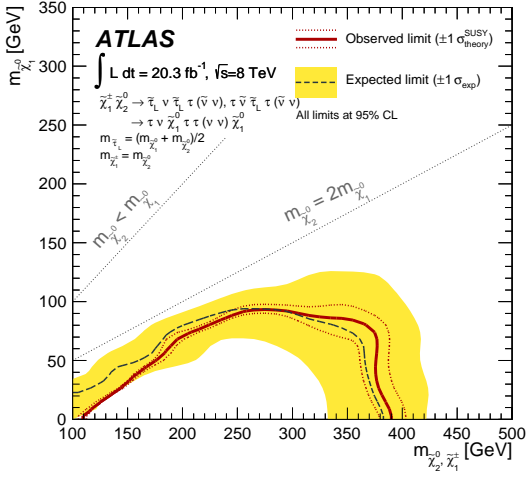}}
\subfigure[$Wh$-mediated simplified model\label{fig:exclLimitsSimpd}]{\includegraphics[width=0.49\textwidth]{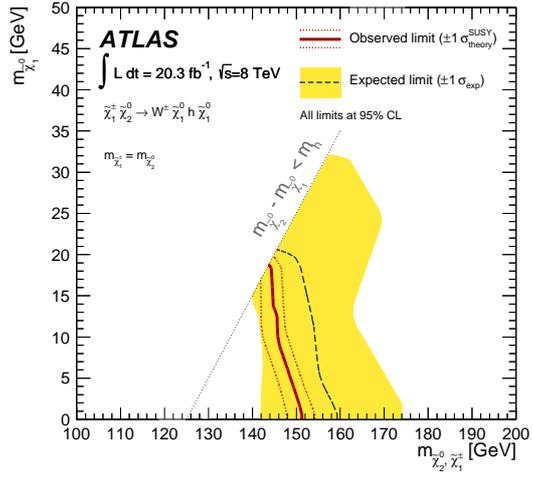}}
\caption{
Observed and expected 95\% CL exclusion contours for chargino and neutralino production in the (a) $\slepL$-mediated, (b) $WZ$-mediated, (c) $\stauL$-mediated and (d) $Wh$-mediated simplified models.
The band around the expected limit shows the $\pm 1 \sigma$ variations of the expected limit, including all uncertainties except theoretical uncertainties on the signal cross-section. 
The dotted lines around the observed limit indicate the sensitivity to $\pm 1 \sigma$ variations of these theoretical uncertainties.
The blue contours in (a) and (b) correspond to the 7$\TeV$ limits from the ATLAS three-lepton analysis~\protect\cite{2012ku}.
Linear interpolation is used to account for the discrete nature of the signal grids.
\label{fig:exclLimitsSimp}}
\end{figure}

In the $\slepL$-mediated simplified model, $\chinoonepm$ and  $\ninotwo$ masses are excluded up to $700 \GeV$ as shown in figure~\ref{fig:exclLimitsSimpa}. 
The region SR0$\tau$a-bin20 offers the best sensitivity to scenarios with high $\chinoonepm$ and $\ninotwo$ masses, and the low-$\msfos$ SR0$\tau$a bins to the small $m_\ninotwo - m_\ninoone$ scenarios. 
In the $WZ$-mediated simplified model shown in figure~\ref{fig:exclLimitsSimpb}, $\chinoonepm$ and $\ninotwo$ masses are excluded up to $345 \GeV$ for massless $\ninoone$.  
The region SR0$\tau$a-bin16 offers the best sensitivity to scenarios with high $\chinoonepm$ and $\ninotwo$ masses, and SR0$\tau$a-bin01 to the small $m_\ninotwo - m_\ninoone$ scenarios. 
The results in the signal regions lead to a weaker (stronger) observed exclusion than expected for the compressed (high-mass $\chinoonepm$, $\ninotwo$) scenarios in both the $\slepL$-mediated and $WZ$-mediated simplified models. 
In the $WZ$-mediated simplified model, there is a reduced sensitivity to scenarios in the $m_{\ninotwo} - m_{\ninoone} = m_Z$ region as the signal populates regions with high $WZ$ background. 
These limits improve those reported by ATLAS in ref.~\cite{2012ku} by $\sim$200$ \GeV$.

In the $\stauL$-mediated simplified model, $\chinoonepm$ and $\ninotwo$ masses are excluded up to $380 \GeV$ for massless $\ninoone$ as shown in figure~\ref{fig:exclLimitsSimpc}.
The low $\msfos$ SR0$\tau$a bins offer the best sensitivity to the small $m_\ninotwo-m_\ninoone$ scenarios, and SR2$\tau$a to the high-mass $\chinoonepm$, $\ninotwo$ scenarios. 
The results in the low $\msfos$ SR0$\tau$a bins lead to a weaker observed exclusion than expected for the compressed scenarios.

In the $Wh$-mediated simplified model shown in figure~\ref{fig:exclLimitsSimpd},  $\chinoonepm$ and $\ninotwo$ masses are excluded up to $148 \GeV$.
The regions SR0$\tau$a, SR0$\tau$b, SR1$\tau$ and SR2$\tau$b offer the best sensitivity in this simplified model when statistically combined. 
The results in some SR0$\tau$a bins and SR1$\tau$ are responsible for the observed exclusion being slightly weaker than expected.

\begin{figure}[h]
\centering
\subfigure[~pMSSM $\sleptonR$, $M_1$=100$ \GeV$\label{fig:exclLimitspMSSMa}]{\includegraphics[width=0.43\textwidth]{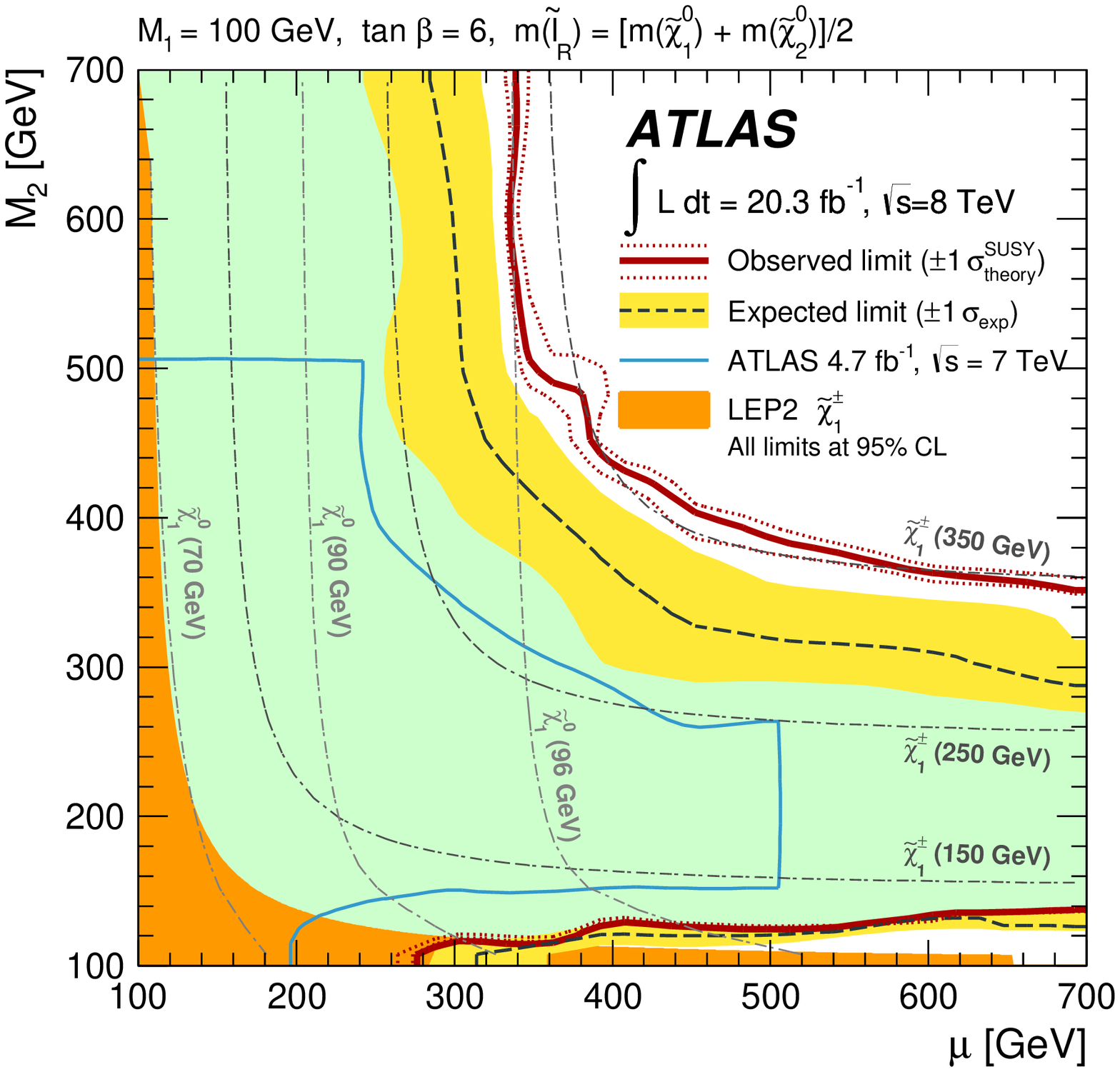}}
\subfigure[~pMSSM $\sleptonR$, $M_1$=140$ \GeV$\label{fig:exclLimitspMSSMb}]{\includegraphics[width=0.43\textwidth]{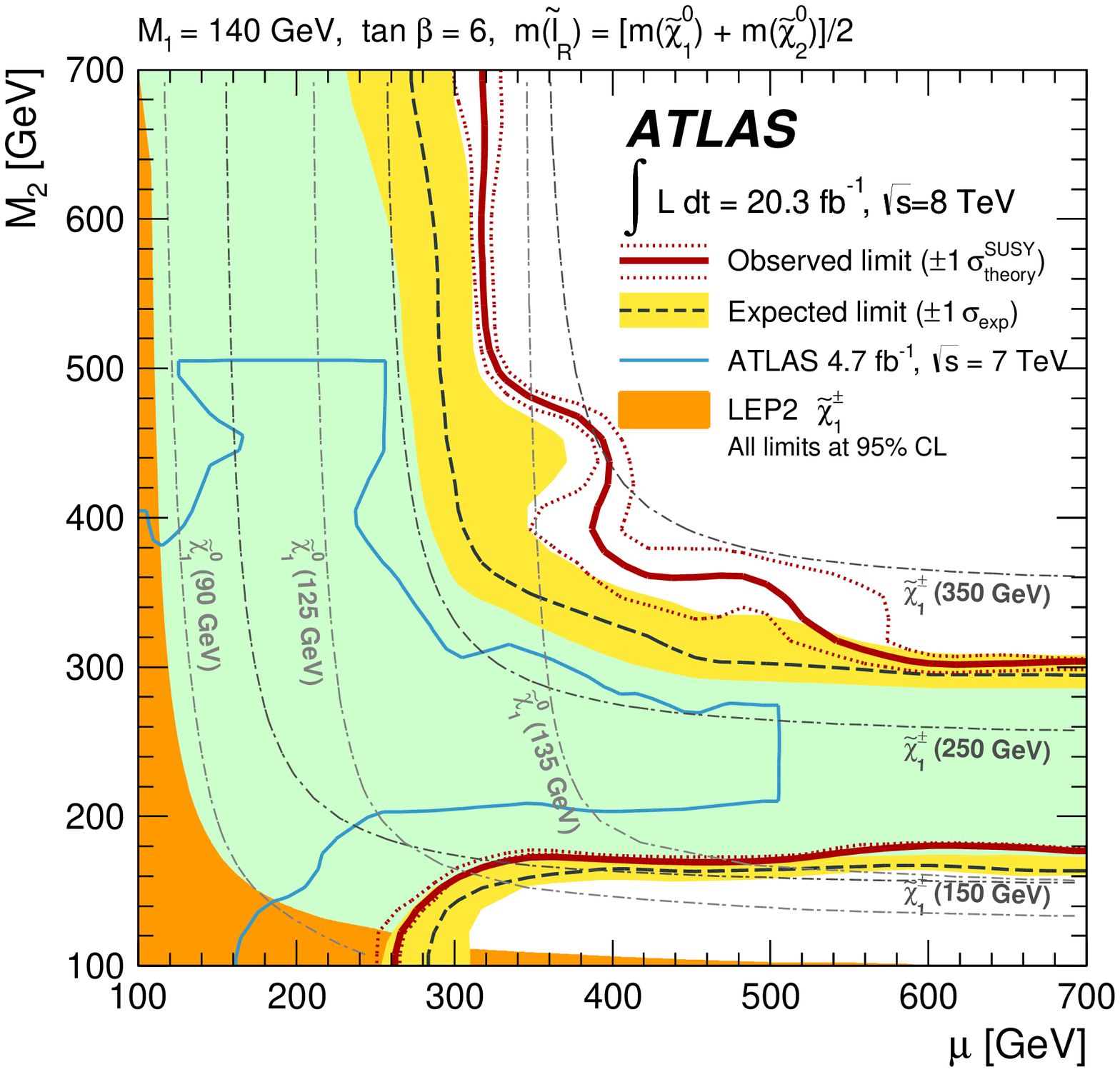}}
\subfigure[~pMSSM $\sleptonR$, $M_1$=250$ \GeV$\label{fig:exclLimitspMSSMc}]{\includegraphics[width=0.43\textwidth]{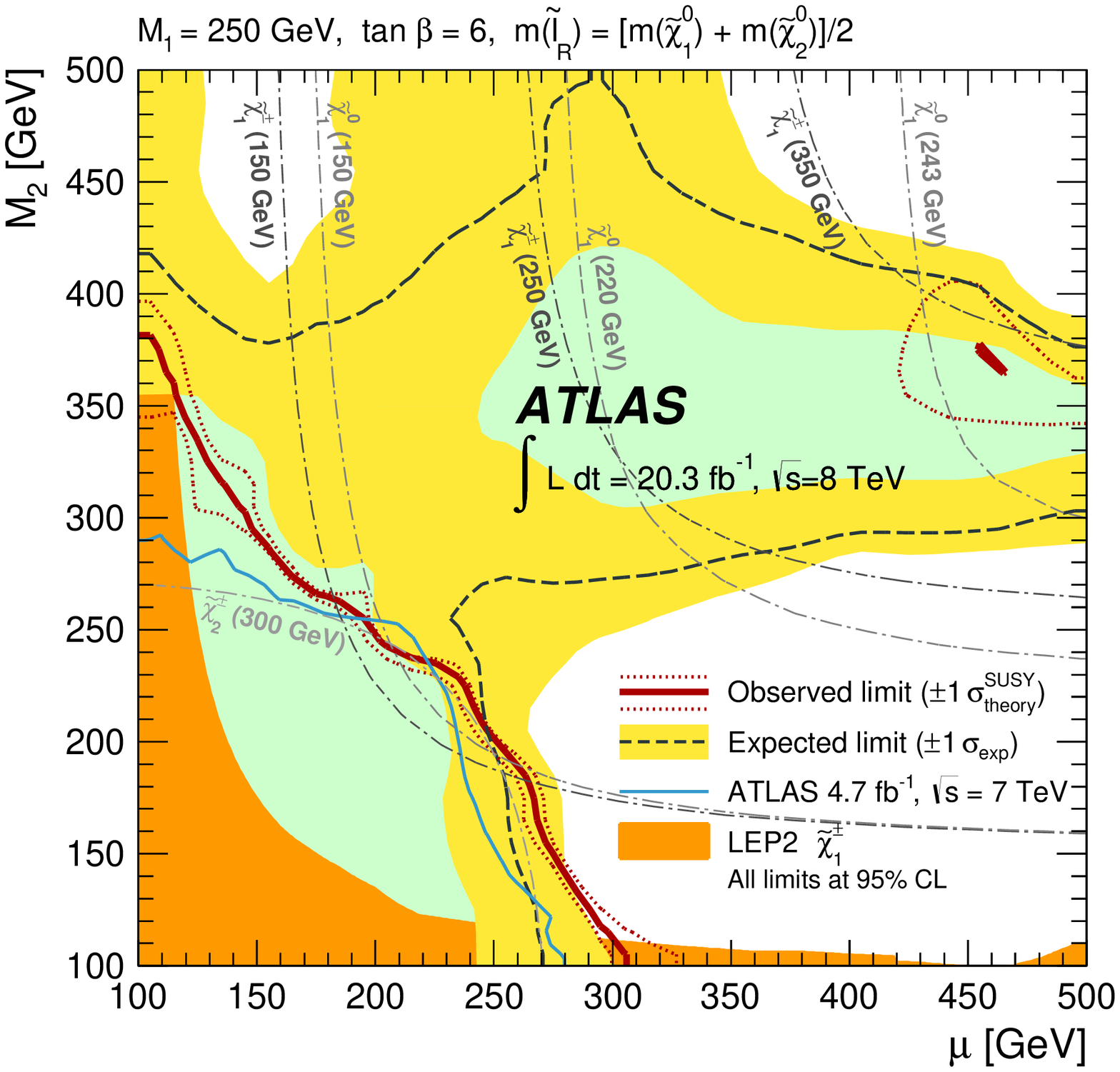}}
\subfigure[~pMSSM $\stauR$, $M_1$=75$ \GeV$\label{fig:exclLimitspMSSMd}]{\includegraphics[width=0.43\textwidth]{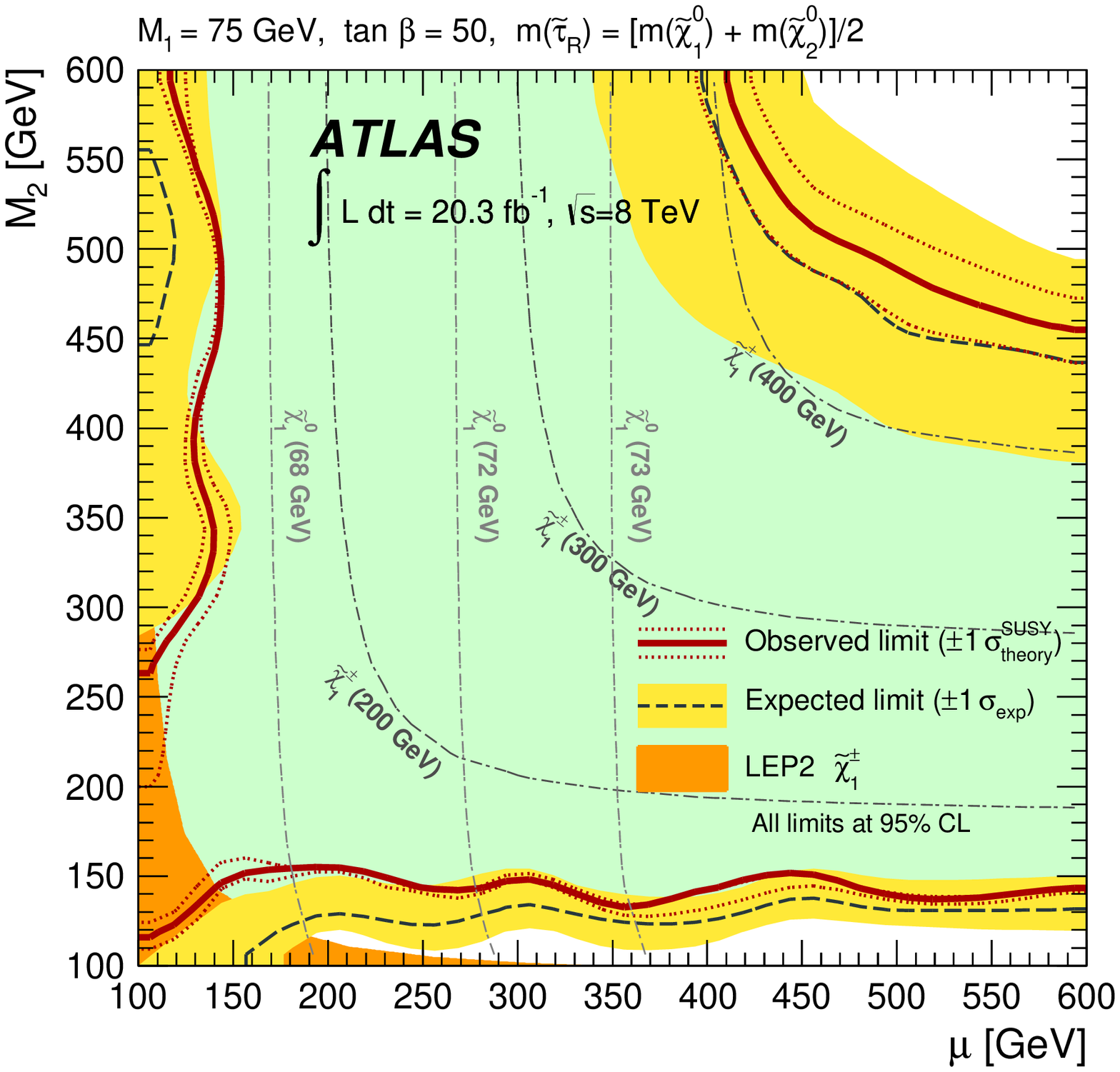}}
\subfigure[~pMSSM no $\slepton$, $M_1$=50$ \GeV$\label{fig:exclLimitspMSSMf}]{\includegraphics[width=0.43\textwidth]{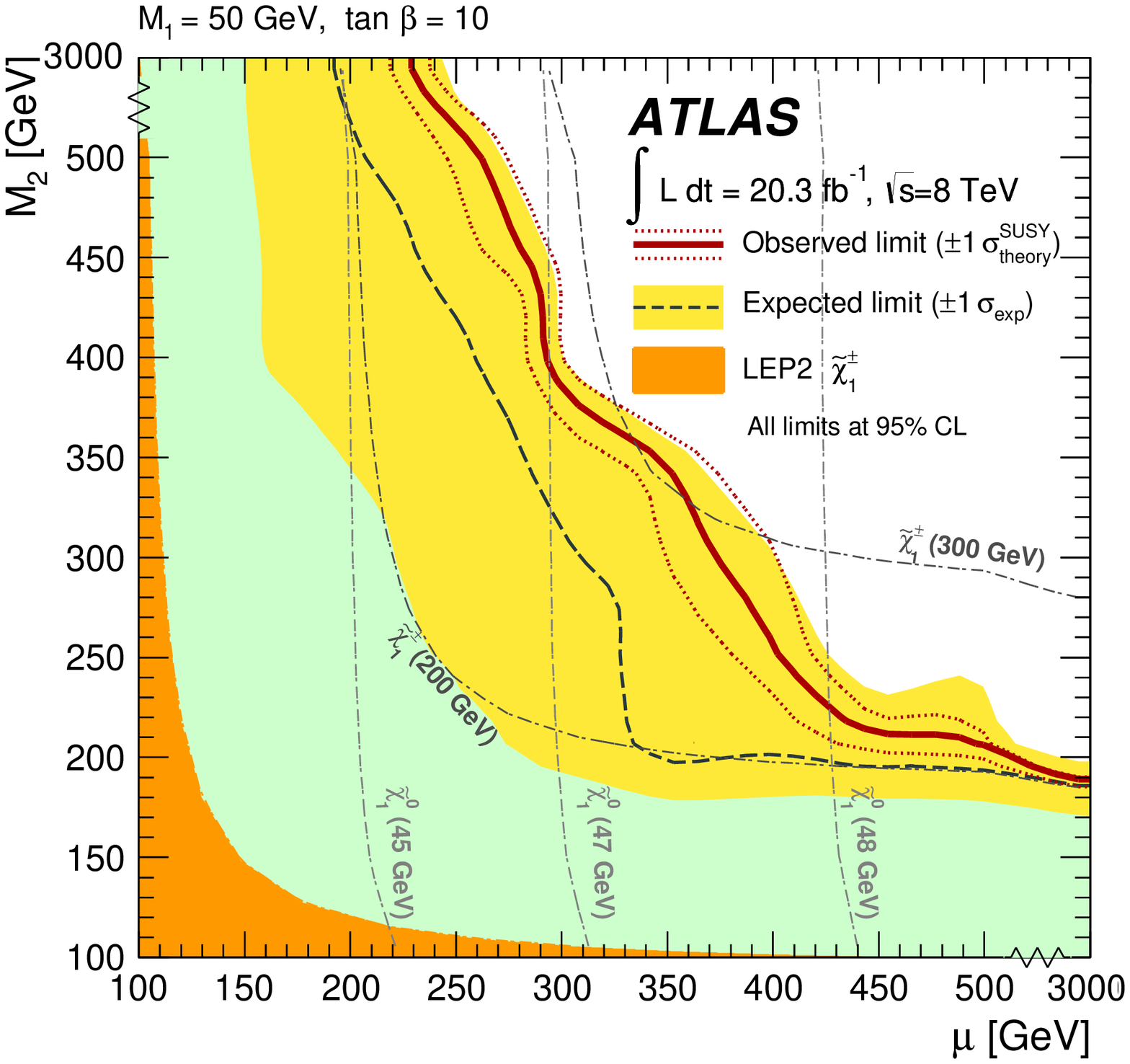}}
\
\caption{
Observed and expected 95\% CL exclusion contours in the pMSSM model with (a)-(c) sleptons, (d) staus and (e) no sleptons.  
The band around the expected limit shows the $\pm 1 \sigma$ variations of the expected limit, including all uncertainties except theoretical uncertainties on the signal cross-section. 
The dotted lines around the observed limit indicate the sensitivity to $\pm 1 \sigma$ variations of these theoretical uncertainties.
The area covered by the $-1\sigma$ expected limit is shown in green.
The blue contours in (a)--(c) correspond to the 7$\TeV$ limits from the ATLAS three-lepton analysis~\protect\cite{2012ku}.
Linear interpolation is used to account for the discrete nature of the signal grids.
\label{fig:exclLimitspMSSM}}
\end{figure}

In the pMSSM scenarios shown in figure~\ref{fig:exclLimitspMSSM}, for a given value of $M_{1}\,$, the sensitivity for high values of $M_{2}\,$ and $\mu$,  and  therefore for high values of  chargino and heavy neutralino masses,  is driven by the decrease of the production cross-section. 
Due to small $m_\ninotwo-m_\ninoone$ or $m_\chinoonepm-m_\ninoone$, limited sensitivity is found  in the regions with $M_{1}\,$$\sim\,$$M_{2}\,$$\ll\,$$\mu$, as seen in figures~\ref{fig:exclLimitspMSSM}(a--d). 
In the case of the pMSSM $\stauR$ scenario, the small mass splittings also reduce the sensitivity in the $M_{1}\sim \mu \ll M_{2}$ region. 
In the pMSSM $\sleptonR$ scenario with $M_1\,$$=\,$250$\GeV$, the $M_2\gtrsim250 \GeV$ and $\mu\gtrsim250 \GeV$ region is characterised by small $m_\chinoonepm-m_\ninoone$, and due to the results in SR0$\tau$a-bin01 the observed exclusion region is significantly smaller than that expected.
For the pMSSM no $\slepton$ scenarios, in the region with $M_2\gtrsim200 \GeV$ and $\mu\gtrsim200 \GeV$ the decay mode $\ninotwo \rightarrow h \ninoone$ is kinematically allowed and reduces the sensitivity.

\section{Conclusions}

This paper describes a search for the production of charginos and neutralinos decaying into final states with three leptons ($e$, $\mu$, $\tau$) and missing transverse momentum. 
The analysis uses \lumi\ of $\sqrt{s}\,$$=\,8\TeV$ proton-proton collision data delivered by the LHC and recorded with the ATLAS detector in 2012. 
No significant excess of events above SM expectations is found in data. 
The null result is interpreted in simplified SUSY models and in various pMSSM scenarios.  
For the simplified SUSY models with intermediate slepton decays, degenerate $\chinoonepm$ and $\ninotwo$ masses up to 700$ \GeV$ are excluded for large mass differences with the $\ninoone$, 
while for the simplified SUSY models with gauge boson decays, the exclusion reaches 345$ \GeV$. 
These limits improve upon the previous ATLAS results in ref.~\cite{2012ku} by $\sim$200$ \GeV$.
For the simplified SUSY models with intermediate staus, degenerate $\chinoonepm$ and $\ninotwo$ masses up to 380$ \GeV$ are excluded, 
while for the simplified SUSY models with intermediate Higgs boson decays, degenerate $\chinoonepm$ and $\ninotwo$ masses up to 148$ \GeV$ are excluded.

\FloatBarrier


\input{Acknowledgements.tex}

\bibliographystyle{JHEP}
\raggedright
\bibliography{paper}

\input{atlas_authlist.tex}


\end{document}

%% file: Acknowledgements.tex



\section{Acknowledgements}

We thank CERN for the very successful operation of the LHC, as well as the
support staff from our institutions without whom ATLAS could not be
operated efficiently.

We acknowledge the support of ANPCyT, Argentina; YerPhI, Armenia; ARC,
Australia; BMWF and FWF, Austria; ANAS, Azerbaijan; SSTC, Belarus; CNPq and FAPESP,
Brazil; NSERC, NRC and CFI, Canada; CERN; CONICYT, Chile; CAS, MOST and NSFC,
China; COLCIENCIAS, Colombia; MSMT CR, MPO CR and VSC CR, Czech Republic;
DNRF, DNSRC and Lundbeck Foundation, Denmark; EPLANET, ERC and NSRF, European Union;
IN2P3-CNRS, CEA-DSM/IRFU, France; GNSF, Georgia; BMBF, DFG, HGF, MPG and AvH
Foundation, Germany; GSRT and NSRF, Greece; ISF, MINERVA, GIF, I-CORE and Benoziyo Center,
Israel; INFN, Italy; MEXT and JSPS, Japan; CNRST, Morocco; FOM and NWO,
Netherlands; BRF and RCN, Norway; MNiSW and NCN, Poland; GRICES and FCT, Portugal; MNE/IFA, Romania; MES of Russia and ROSATOM, Russian Federation; JINR; MSTD,
Serbia; MSSR, Slovakia; ARRS and MIZ\v{S}, Slovenia; DST/NRF, South Africa;
MINECO, Spain; SRC and Wallenberg Foundation, Sweden; SER, SNSF and Cantons of
Bern and Geneva, Switzerland; NSC, Taiwan; TAEK, Turkey; STFC, the Royal
Society and Leverhulme Trust, United Kingdom; DOE and NSF, United States of
America.

The crucial computing support from all WLCG partners is acknowledged
gratefully, in particular from CERN and the ATLAS Tier-1 facilities at
TRIUMF (Canada), NDGF (Denmark, Norway, Sweden), CC-IN2P3 (France),
KIT/GridKA (Germany), INFN-CNAF (Italy), NL-T1 (Netherlands), PIC (Spain),
ASGC (Taiwan), RAL (UK) and BNL (USA) and in the Tier-2 facilities
worldwide.

%% file: atlas_authlist.tex
\begin{flushleft}
{\Large The ATLAS Collaboration}

\bigskip

G.~Aad$^{\rm 84}$,
T.~Abajyan$^{\rm 21}$,
B.~Abbott$^{\rm 112}$,
J.~Abdallah$^{\rm 152}$,
S.~Abdel~Khalek$^{\rm 116}$,
O.~Abdinov$^{\rm 11}$,
R.~Aben$^{\rm 106}$,
B.~Abi$^{\rm 113}$,
M.~Abolins$^{\rm 89}$,
O.S.~AbouZeid$^{\rm 159}$
% ...
\bigskip
\\
% ...
$^{11}$ Institute of Physics, Azerbaijan Academy of Sciences, Baku, Azerbaijan\\
% ...
$^{21}$ Physikalisches Institut, University of Bonn, Bonn, Germany\\
% ...
$^{84}$ CPPM, Aix-Marseille Universit{\'e} and CNRS/IN2P3, Marseille, France\\
% ...
$^{89}$ Department of Physics and Astronomy, Michigan State University, East Lansing MI, United States of America\\
% ...
$^{106}$ Nikhef National Institute for Subatomic Physics and University of Amsterdam, Amsterdam, Netherlands\\
% ...
$^{112}$ Homer L. Dodge Department of Physics and Astronomy, University of Oklahoma, Norman OK, United States of America\\
$^{113}$ Department of Physics, Oklahoma State University, Stillwater OK, United States of America\\
% ...
$^{116}$ LAL, Universit{\'e} Paris-Sud and CNRS/IN2P3, Orsay, France\\
% ...
$^{152}$ Institute of Physics, Academia Sinica, Taipei, Taiwan\\
% ...
$^{159}$ Department of Physics, University of Toronto, Toronto ON, Canada\\
% ...
\end{flushleft}
